\documentclass[aps,prl,reprint,groupedaddress,showpacs,preprintnumbers,amsmath,amssymb,superscriptaddress,nofootinbib,onecolumn,notitlepage]{revtex4-1}

\usepackage{flushend}

\usepackage{graphicx}
\usepackage[export]{adjustbox}
\usepackage{bm}

\usepackage{enumerate}
\usepackage{subfigure}
\usepackage[arrow]{hhtensor}     
\usepackage{leftidx}

\newcommand{\me}{\mathrm{e}}
\newcommand{\mi}{\mathrm{i}}
\newcommand{\dif}{\mathrm{d}}

\newcommand\blfootnote[1]{
  \begingroup
  \renewcommand\thefootnote{}\footnote{#1}%
  \addtocounter{footnote}{-1}%
  \endgroup
}

\DeclareMathAlphabet{\mathcal}{OMS}{cmsy}{m}{n}

\allowdisplaybreaks[4]

\begin{document}


\title{Time Circular Birefringence in Time-Dependent Magnetoelectric Media}

\author{Ruo-Yang Zhang}
\affiliation{
 Theoretical Physics Division, Chern Institute of Mathematics, Nankai University, Tianjin,  300071, China
}
\affiliation{
Department of Physics, The Hong Kong University of Science and Technology, Clear Water Bay, Hong Kong, China
}

\author{Yan-Wang Zhai}%
\affiliation{
School of Physics, Beijing Institute of Technology, Beijing, 100081, China
}

\author{Shi-Rong Lin}%
\affiliation{
School of Physics, Beijing Institute of Technology, Beijing, 100081, China
}

\author{Qing Zhao}%
\affiliation{
School of Physics, Beijing Institute of Technology, Beijing, 100081, China
}

\author{Weijia Wen}
\affiliation{
Department of Physics, The Hong Kong University of Science and Technology, Clear Water Bay, Hong Kong, China
}

\author{Mo-Lin Ge}%
\affiliation{
 Theoretical Physics Division, Chern Institute of Mathematics, Nankai University, Tianjin,  300071, China
}
\affiliation{
School of Physics, Beijing Institute of Technology, Beijing, 100081, China
}


\date{\today}
\maketitle
\blfootnote{Correspondence and requests for materials should be addressed to R.-Y. Z.(zhangruoyang@gmail.com), Q. Z.(qzhaoyuping@bit.edu.cn) or M.-L. G.(geml@nankai.edu.cn)}

\textbf{
    Light traveling in time-dependent media has many extraordinary properties which can be utilized to convert frequency, achieve temporal cloaking, and simulate cosmological phenomena. In this paper, we focus on time-dependent axion-type magnetoelectric (ME) media, and prove that light in these media always has two degenerate modes with opposite circular polarizations corresponding to one wave vector $\bm{k}$, and name this effect ``time circular birefringence'' (TCB).  By interchanging the status of space and time, the pair of TCB modes can appear simultaneously via ``time refraction'' and ``time reflection'' of a linear polarized incident wave at a time interface of ME media. The superposition of the two TCB modes causes the ``time Faraday effect'', namely the globally unified polarization axes rotate with time. A circularly polarized Gaussian pulse traversing a time interface is also studied. If the wave-vector spectrum of a pulse mainly concentrates in the non-traveling-wave band, the pulse will be trapped with nearly fixed center while its intensity will grow rapidly. In addition, we propose an experimental scheme of using molecular fluid with external time-varying electric and magnetic fields both parallel to the direction of light to realize these phenomena in practice.}



\maketitle 

\vspace{20pt}


\noindent
For general linear nondispersive bianisotropic media, the constitutive relations are~
\begin{equation}\label{constitutive relation}
  \bm{D}=\matr{\varepsilon}\cdot\bm{E}+\matr{\xi}\cdot\bm{B},\quad\ \bm{H}=-\matr{\zeta}\cdot\bm{E}+\matr{\mu}^{-1}\cdot\bm{B}.
\end{equation}
The tensors $\matr{\xi},\ \matr{\zeta}$ correspond to the magnetoelectric (ME) cross polarizations. A ME medium satisfying ${\xi^i}_j=-{\zeta_j}^i$ is reciprocal, {\it e.g.} chiral medium, otherwise it is nonreciprocal. The nonreciprocal ME effect was first discovered in $\rm Cr_2O_3$~\cite{dzyaloshinskii1960magneto,folen1961anisotropy,landau1984electrodynamics}, and has attracted wide attention both in condensed matter physics~\cite{fiebig2005revival,eerenstein2006multiferroic,pyatakov2012magnetoelectric,qi2008topological,li2010dynamical,qi2009inducing,essin2009magnetoelectric,nomura2011surface} and in optics~\cite{de2001post,hehl2005linear,obukhov2005measuring,jung2004optical,kida2006enhanced,kida2007optical,takahashi2012magnetoelectric,kamenetskii2009tellegen,tse2010giant,tse2010magneto,tse2011magneto,Bliokh2014magnetoelectric}. It has been shown that a nonreciprocal ME medium with nonzero $\mathrm{Tr}({\xi^i}_j+{\zeta_j}^i)$ can separate a real term $\Theta$ from the ME coupling~\cite{de2001post,hehl2005linear,obukhov2005measuring}.
If we are only concerned with this term, the two ME coefficients reduce to isotropy: ${\xi^i}_j={\zeta_j}^i=\Theta\,\delta^i_j$. Then the Maxwell equations can be expressed as the axion-like form~\cite{wilczek1987two,qi2008topological}
with the virtual electric displacement
$\bm{\widetilde{D}}= \matr{\varepsilon}\cdot\bm{E}$ and the virtual magnetic field $\bm{\widetilde{H}}= \matr{\mu}^{-1}\cdot\bm{B}$
excluding the electric and magnetic cross polarizations. By redefining a virtual excitation tensor $\widetilde{G}^{\mu\nu}=G^{\mu\nu}+\Theta\, \leftidx{^\star}F^{\mu\nu}$ constructed from the virtual fields: $\widetilde{G}^{0i}=-c\widetilde{D}^i$, $\widetilde{G}^{ij}=-\epsilon^{ijk}\widetilde{H}_k$, the lagrangian density 
in the isotropic ME media can be written as same as the one in axion electrodynamics~\cite{qi2008topological,li2010dynamical,qi2009inducing}:
\begin{equation}\label{lagrangian}
  \mathcal{L}=-\frac{1}{4c}\widetilde{G}^{\mu\nu}F_{\mu\nu}+\frac{1}{c}A_\mu J^\mu+\frac{1}{4c}\Theta F_{\mu\nu}\,\leftidx{^\star}F^{\mu\nu},
\end{equation}
where $\leftidx{^\star}F^{\mu\nu}=\frac{1}{2}\epsilon^{\mu\nu\alpha\beta}F_{\alpha\beta}$ is the Hodge dual of $F_{\alpha\beta}$.
In Eq.~(\ref{lagrangian}), the last term $\mathcal{L}_{\Theta}=\frac{1}{4c}\Theta F_{\mu\nu}\,\leftidx{^\star}F^{\mu\nu}=\Theta\, \bm{E}\cdot\bm{B}$  just corresponds to the axion coupling, and $\Theta(x^\mu)$ corresponds to the axion field. 
Correspondingly, the 4-D Maxwell equation also holds the axion-like form
$\partial_\mu\widetilde{G}^{\mu\nu}=J^{\nu}+\partial_{\mu}\Theta\,\leftidx{^\star}F^{\mu\nu}$.
Since $\bm{E}$ is a polar vector while $\bm{B}$ is an axial vector, $\Theta$ must be a pseudoscalar to guarantee that  the lagrangian density is a Lorentz scalar.

Axion was originally proposed as a hypothetical elementary particle~\cite{peccei1977cp}, while it won great interests in condensed matter physics recently because of the significant discovery that an effective quantized axion field
can be induced in topological insulators when time reversal symmetry is weakly broken~\cite{qi2008topological,li2010dynamical,qi2009inducing,essin2009magnetoelectric,nomura2011surface}.
Actually, since $\Theta$ is a pseudoscalar, the axion-type ME coupling only exists in the systems where both the  time reversal ($T$) and the parity ($P$) symmetries are broken but the combined $PT$ symmetry is held~\cite{landau1984electrodynamics}.
There is no visible effect for light traveling in globally constant axion field, however, a Kerr or Faraday rotation can be detected for lights reflected or refracted by the surface of an axion medium~\cite{qi2008topological,tse2010giant,tse2010magneto,tse2011magneto},
which essentially originates from the sudden change of $\Theta$ at the spatial interface~\cite{qi2008topological}. Noteworthy, 
a type of circular birefringence, known as Carroll-Field-Jackiw (CFJ) birefringence, can emerge in Chern-Simons modified electrodynamics~\cite{carroll1990limits}. And Y. Itin proved that the CFJ birefringence can be alternatively caused by a space-and-time-dependent axion field in geometric optics approximation~\cite{itin2004carroll,itin2008wave}. The CFJ birefringence is generally anisotropic in space, whereas it reduces to isotropy when the 4-gradient $\partial_\mu\Theta$ is timelike, {\it i.e.} the axion field only changes with time.

Light traveling in time-dependent media has many extraordinary properties which can be utilized to achieve frequency conversion~\cite{ginis2010frequency,cummer2011frequency}, temporal cloaking~\cite{mccall2011spacetime,fridman2012demonstration,lukens2013temporal,chremmos2014temporal}, and to simulate cosmological phenomena~\cite{philbin2008fiber,westerberg2014experiment} {\it etc}. In this paper, we focus on time-dependent axion-type ME media, and prove that light in these media always has two oppositely circularly polarized modes corresponding to one wave vector $\bm{k}$ but not limited to geometric optics approximation.  The key idea of this paper is to interchange the status of space and time. We will show that the pair of TCB modes can appear simultaneously via the ``time refraction'' and ``time reflection'' of a linearly polarized incident wave at a time-discontinuous interface of the ME media. The superposition of two TCB modes causes the “time Faraday effect” which is a novel effect as a temporal counterpart of the ordinary spatial Faraday effect or optical activity. Further discussions about the propagating velocities of energy and information for TCB modes and about the time refraction and reflection of Gaussian pulse at time interfaces in ME media are also provided. Furthermore, we put forward an experimental scheme to generate the effective time-dependent axion-type ME media controlled by time-varying external electric field $\mathcal{E}$ and magnetic field $\mathcal{B}$ parallel to each other which offers a practical way to realize the novel phenomena predicted in this paper.

\vspace{20pt}
\noindent
\textbf{\Large Results}\\[3pt]
\noindent\textbf{\large Time circular birefringence and time Faraday effect.\ }
In time-dependent axion-type ME media, the magnetic induction obeys the wave equation
\begin{equation}\label{wave equation}
  \nabla^2\bm{B}+\mu\dot{\Theta}\nabla\times\bm{B}-\mu\frac{\partial}{\partial t}\left(\varepsilon\frac{\partial\bm{B}}{\partial t}\right)=0,
\end{equation}
where the dot over $\Theta$ denotes the derivative with respect to time, and $\varepsilon,\ \mu,\ \Theta$ are all functions of time in general. While the $P$ and $T$ symmetries are both broken in Eq.~(\ref{wave equation}), the combined $PT$ symmetry is preserved. Considering the class of solutions $\bm{B}=\bm{T}(t)\,\me^{\mi\bm{k}\cdot\bm{r}}$ with a constant wave vector $\bm{k}$, the temporal part satisfies $\bm{k}\cdot\bm{T}(t)=0$ due to $\nabla\cdot\bm{B}=0$. Therefore, the temporal part can be further separated into two independent circularly polarized portions $\bm{T}(t)=T_-\bm{\hat{U}}_+ +T_+\bm{\hat{U}}_-$ obeying the following equations respectively
\begin{equation}\label{eigenequation}
  \frac{\dif^2T_\pm}{\dif t^2}+\frac{\dif \ln\varepsilon}{\dif t}\frac{\dif T_\pm}{\dif t}+v^2 k(k\pm\mu\dot{\Theta})T_\pm=0,
\end{equation}
where $\bm{\hat{U}}_\pm=\frac{1}{\sqrt{2}}(\bm{\hat{x}}\pm\mi\bm{\hat{y}})$ are the circularly polarized bases with choosing the direction of $\bm{k}$ to be $z$ axis, and $v^2=1/\varepsilon\mu$.
As a result, there always exists a pair of circularly birefringent modes $T_\pm$ for a given wave vector $\bm{k}$ in time-dependent axion-type media:
$
  \bm{B}_{\pm}= T_\pm(t)\me^{\mi\bm{k}\cdot\bm{r}} \bm{\hat{U}}_\mp.
$
We call this effect the time circular birefringence (TCB). If $\dot{\Theta}=0$, the two distinct equations of $T_\pm$ reduce to an identical one, and the birefringent phenomenon vanishes. Thereby TCB is entirely induced by the time varying axion field. In addition, TCB happens in isotropic media, thus it is different from both the ordinary birefringence in uniaxial or biaxial crystals and the ME Jones birefringence \cite{graham1983jones,roth2000observation,roth2000Magneto-electric,rizzo2003jones} which are all caused by the anisotropy of materials.  TCB is also different from the optical active circular birefringence (OACB), 
because TCB is generated from the temporal nonhomogeneity of the nonreciprocal ME media but OACB is a reciprocal magnetoelectric effect originating from the chirality of molecules.

\begin{figure}[t]
\includegraphics[width=0.7\columnwidth,clip]{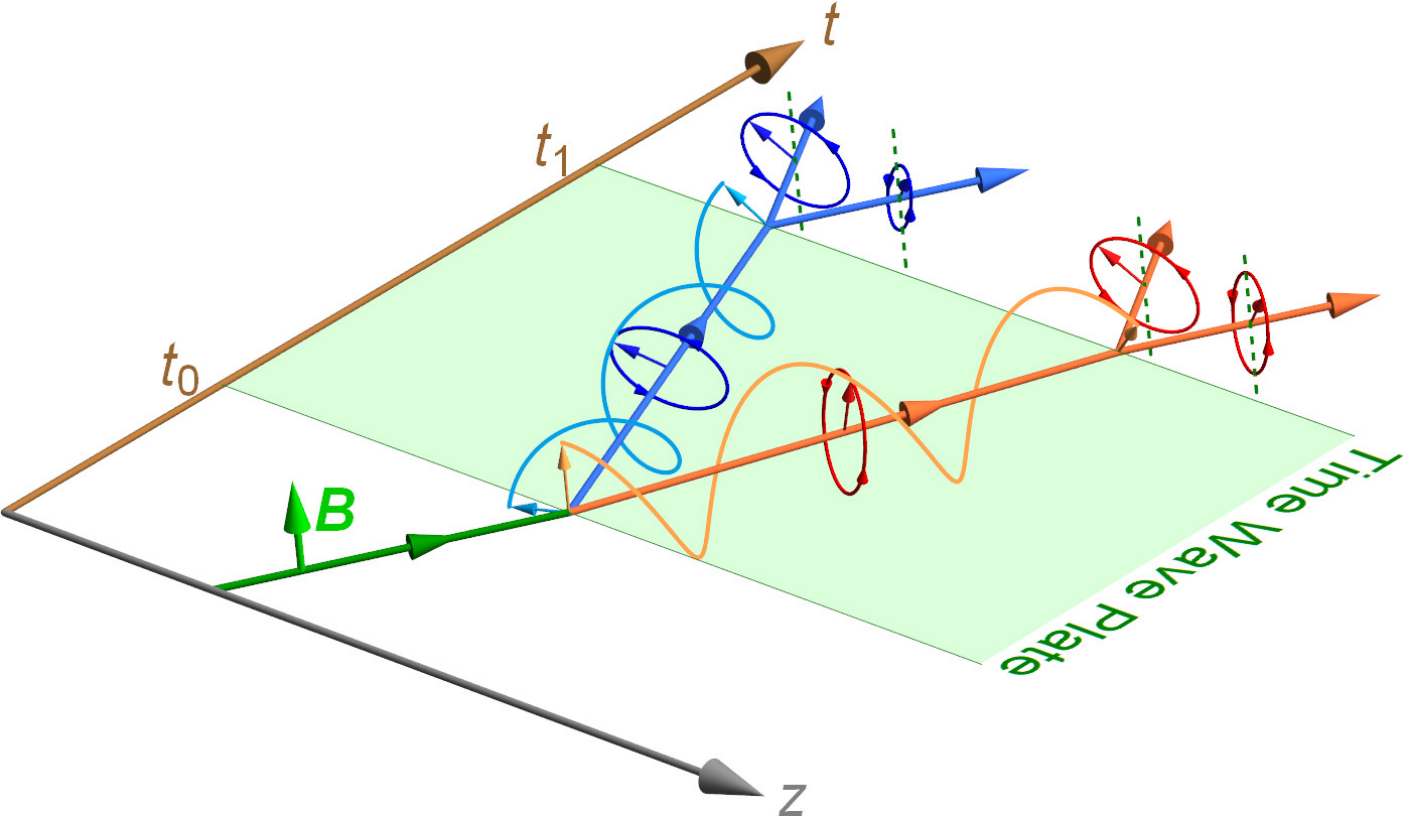}
\caption{\label{fig1}Illustration of time refraction, time reflection and time Faraday rotation for a linearly polarized light incident upon a time wave plate with time-dependent ME coefficient $\Theta=\beta \,t/\mu_1\ (t_0<t<t_1)$. At $t_0$, the wave splits into a time refracted part and a time reflected part. The two parts are both elliptically polarized, but their major axes rotate with time. After the second interface $t_1$, the polarization axes of the four outgoing waves have angular differences with respect to the polarization of the incident wave.}
\end{figure}

For traditional birefringent effects, two different wave numbers $k$ correspond to one frequency. One can realize the two birefringent states just via shooting a beam onto a birefringent medium subject to the temporal-phase-matching condition $\me^{-\mi\omega t}=\me^{-\mi\omega' t}$ at the spatial interface. However, the temporal parts $T_\pm(t)$ of the pair of TCB modes corresponding to a fixed wave number are different, and accordingly could not match the temporal phase factor of the incident wave simultaneously. This difference gives rise to a handicap for realizing this pair of circular polarized states in practice. To overcome this difficulty, we think up the idea of ``time discontinuous media'' by analogy with the ``spatial discontinuous media'' used in traditional birefringent systems, then the spatial phase factor, $\me^{\mi \bm{k}\cdot\bm{r}}$, should be matched at time interfaces. Considering a time-dependent medium $\varepsilon(t),\ \mu(t),\ \Theta(t)$ discontinuous at a time interface $t_0$, we can get the temporal boundary conditions of electromagnetic fields by integrating Maxwell equations over an infinitesimal time interval across $t_0$ \cite{mendonca2002time,xiao2014reflection}:
\begin{equation}\label{boundary conditions}
  \left.\bm{D}\right|_{t_0-}=\left.\bm{D}\right|_{t_0+},\quad \left.\bm{B}\right|_{t_0-}=\left.\bm{B}\right|_{t_0+},
\end{equation}
while $\bm{E}$ and $\bm{H}$ are generically discontinuous at the time interface.

Just as spatial optical wave plate devices, we analyze light propagating in a ``time wave plate'' with piecewise  medium parameters: $\varepsilon_0,\ \mu_0,\ \Theta_0$ are constant when $t<t_0$; $\varepsilon_1(t)$, $\mu_1(t)$, $\Theta_1(t)$ are some continuous functions when $t_0<t<t_1$; $\varepsilon_2$, $\mu_2$, $\Theta_2$ are also constant when $t>t_1$, as shown in Fig.~\ref{fig1}. For a linearly polarized incident wave $\bm{B}^{\rm in}=\bm{A}\me^{\mi(kz-\omega_0t)}$ with $\omega_0=k/\sqrt{\varepsilon_0\mu_0}$ and $\bm{A}=\sum_\pm (A/\sqrt{2})\me^{\mp\mi\phi}\hat{\bm{U}}_\pm$, the wave will become the sum of the two TCB modes $\bm{B}=\sum_\pm \bm{B}_\pm$ at $t_0$. Moreover, there always exist two linearly independent solutions for Eq.~(\ref{eigenequation}) which are complex conjugates of each other: $T^1_\pm(t)=T^2_\pm(t)^*$, then the general solution of Eq.~(\ref{eigenequation}) is their superposition: $T_\pm(t)=A_\pm^1T_\pm^1(t)+A_\pm^2T_\pm^2(t)$ , and the two TCB states can be further separated as $\bm{B}_\pm=A_\pm^1\bm{B}_\pm^1+A_\pm^2\bm{B}_\pm^2$. It can be proved that the momentums of the two branches $\bm{B}_\pm^1$ and $\bm{B}_\pm^2$ are always in opposite directions, i.e. one branch always propagates along the incident direction (for convenience, let it be $\bm{B}_\pm^1$), while the other (let it be $\bm{B}_\pm^2$) is always along the opposite. As a result, $\bm{B}_\pm^1$ and $\bm{B}_\pm^2$ are exactly the ``time refraction'' and  ``time reflection'' of the corresponding TCB modes at the time interface $t_0$ (see the supplementary information for more discussions).


A simplified case is $\dot{\Theta}_1(t)\equiv \beta/\mu_1>0$, and $\varepsilon_1,\ \mu_1$ are both constant. Then the TCB modes are identical with the CFJ modes obtained in geometric optics approximation~\cite{carroll1990limits,itin2004carroll,itin2008wave}, therefore the\ light splits into two plane waves 
\begin{equation}\label{time reflection and refraction2}
  \bm{B}^\sigma=\sum_\pm A_\pm^\sigma\,\me^{\mi[kz+\delta^\sigma\omega_\pm t']}\bm{\hat{U}}_\mp\quad (\sigma=1,2),
\end{equation}
as $t'=t-t_0>0$. The dispersion relations of two TCB modes are $\omega_\pm=v_1\,k\sqrt{(k\pm\beta)/k}$, and the coefficients determined by the temporal boundary conditions are
\begin{equation}\label{coefficent}
  A_\pm^\sigma=\frac{A\me^{\mi\left(\pm\phi-\omega_0 t_0\right)}}{2\sqrt{2}}\left[1-\delta^{\sigma}\frac{\varepsilon_0\omega_0\mp\mi k[\Theta_1(t_0)-\Theta_0]}{\varepsilon_1\omega_\pm}\right],
\end{equation}
with $\delta^\sigma=(-1)^\sigma$.  According to the dispersion relations, the two TCB modes $\bm{B}_\pm$ both have a forbidden band of $k$ for traveling waves: $\pm k\in [-\beta,0]$. Outside the forbidden band,  $\bm{B}^1$ travels along the incident direction, i.e. it is the time refraction, and $\bm{B}^2$ is the time reversal of $\bm{B}^1$. However, a wave should not propagate backwards through time. The practical observable is its real part which propagates opposite to the incident direction in space, therefore, $\bm{B}^2$ is actually the time reflection.
Without loss of the physical generality, a further simplification will applied in the following: $\varepsilon_0=\varepsilon_1=\varepsilon_2$, $\mu_0=\mu_1=\mu_2$, $\Theta_1(t_0)=\Theta_0$, and $\Theta_1(t_1)=\Theta_2$, {\it i.e.} the medium is continuous at $t_0$ and $t_1$ but $\dot\Theta$ is still discontinuous.

\begin{figure}[t]
\includegraphics[width=0.8\columnwidth,clip]{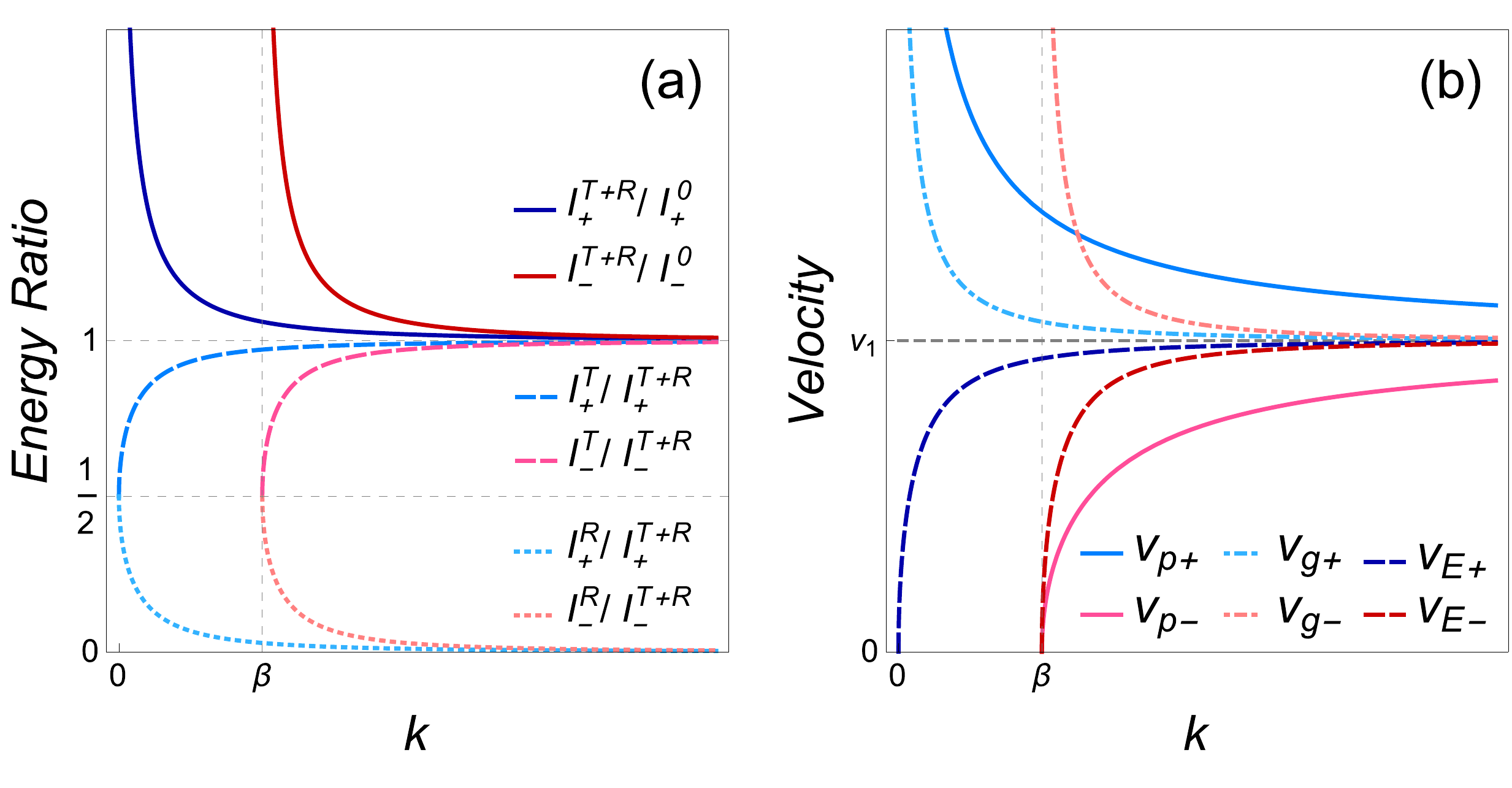}
\caption{\label{fig_velocity}(a) Ratios of total light intensity of refraction and transmission to the incident light intensity $I^{T+R}_\pm/I^0_\pm$, modified transmissivity $I^{T}_\pm/I^{T+R}_\pm$, and modified reflectivity $I^{R}_\pm/I^{T+R}_\pm$ corresponding to the two TCB modes respectively are shown as functions of wave number $k$ (see the supplementary information for more details). (b) Phase velocities $v_{\rm p\pm}$, group velocities $v_{\rm g\pm}$, energy transport velocities $v_{\rm E\pm}$ and front velocity $v_{\rm f}\equiv v_1$ of the two TCB modes versus $k$.}
\end{figure}

The time dependence of media destructs the symmetry of time translation, therefore, the energy of the electromagnetic field is not conserved in general. On the other hand, the lagrangian of time dependent media shown in Eq.~(\ref{lagrangian}) is invariant under spatial translation, so the apparent electromagnetic momentum $\bm{P}=\mathfrak{Re}(\bm{D})\times\mathfrak{Re}(\bm{B})$ must be conserved. Typically, the energy of incident wave does not equal to the total energy of the time refracted and reflected waves at the time interfaces of a time wave plate (see Fig.~\ref{fig_velocity}\,(a)\,), whereas the incident apparent momentum equals to the resultant momentum of the reflected and refracted waves: $\bm{P}^{\rm in}=\bm{P}^1+\bm{P}^2$ (see the supplementary information for general proof). From a photonic point of view, the nonconservation of energy indicates $\omega_0\neq\omega_\pm$, while the conservation of momentum insures $\bm{k}^{\rm in}=\bm{k}^1=\bm{k}^2$ at time interfaces. This fact is different from the case of ordinary refraction and reflection at a spatial interface of two media, in which the energy is conserved, but the normal momentum to the spatial interface isn't conserved because the discontinuity of the media breaks the symmetry of spatial translation.

As shown in Eq.~(\ref{time reflection and refraction2}), the refracted and reflected waves both have two circularly polarized components with different frequencies $\omega_\pm$. The superposition of the two components gives rise to the time Faraday rotation (TFR), namely, the refracted and reflected waves can be rewritten as a sole polarized wave respectively
\begin{equation}\label{time Faraday rotation}
\begin{split}
  \bm{B}^\sigma=&\frac{A}{2}\Big[\left(1-\delta^\sigma\frac{\omega_0\bar\omega}{\omega_+\omega_-}\right)\bm{\hat{x}}^\sigma(t')+\mi\frac{\omega_0\Delta\omega}{\omega_+\omega_-}\bm{\hat{y}}^\sigma(t')\Big]
  \cdot\me^{\mi(kz-\bar\omega\, t'-\omega_0t_0)},\qquad (\sigma=1,2),
\end{split}
\end{equation}
with the time dependent bases
\begin{equation}\label{time dependent bases}
  \begin{pmatrix} \bm{\hat{x}}^\sigma(t') \\ \bm{\hat{y}}^\sigma(t')\end{pmatrix}=
  \begin{pmatrix}
    \cos(\Delta\omega t'+\phi^\sigma)  & -\sin(\Delta\omega t'+\phi^\sigma)\\
    \sin(\Delta\omega t'+\phi^\sigma)  & \cos(\Delta\omega t'+\phi^\sigma)
  \end{pmatrix}
  \begin{pmatrix} \bm{\hat{x}} \\ -\delta^\sigma \bm{\hat{y}}\end{pmatrix},
\end{equation}
where $\bar\omega=(\omega_+ +\omega_-)/2$, $\Delta\omega=(\omega_+-\omega_-)/2$, $\phi^\sigma=(-1)^\sigma\phi$. So both the time refracted and reflected waves can be regarded as  generic elliptically polarized plane waves propagating with the frequency $\bar\omega$, but their polarization ellipses rotate with angular velocity $\Delta\omega$, {\it i.e.} the TFR. Because of the $PT$ symmetry, the refracted and reflected waves rotate in same chirality with respect to their respective propagating directions.  Unlike ordinary magneto-optical Faraday effect or optical activity which both refer to the polarization of a wave changing circularly in its propagating direction, the TFR wave has a unique polarization in the whole space at any fixed time point, however, the polarization rotates with time. Note that the Faraday effect caused by two opposite circularly polarized CFJ waves was also discussed in Ref.~\cite{carroll1990limits}. However, their effect is still a spatial Faraday rotation, {\it i.e.} the two superposed CFJ waves have same frequency $\omega$ but different $k$ and the rotating angle changes with traveling distance,  therefore the TFR caused by the time refraction and time reflection is entirely a novel effect distinct form their discussion.

At the second time interface $t_1$ of the time wave plate, the secondary time refraction and reflection occur. Then the beam will split into four elliptically polarized branches, 
all of which can be written by
\begin{equation}\label{secondary time reflection and refraction}
  \bm{B}^{\sigma\tau}=\frac{A}{2}\Big[a^{\sigma\tau}\bm{\hat{x}}^\sigma(t'_1)+\mi\, b^{\sigma\tau}\bm{\hat{y}}^\sigma(t'_1)\Big]\me^{\mi[kz+\delta^\tau\omega_2\,t'+\varphi_0^\sigma]},
\end{equation}
where $t'_1=t_1-t_0$, and $\varphi_0^\sigma=\delta^\sigma\bar\omega\, t'_1-\omega_0t_0$. The superscript $\sigma\tau$ distinguishes the four branches: $\sigma\tau=11,\ 12$ denote, respectively, the secondary refraction and reflection of the first refracted wave, and $\sigma\tau=22,\ 21$ denote the secondary refraction and reflection of the first reflected wave respectively. Eq.~(\ref{secondary time reflection and refraction}) shows that the Faraday rotating angle of the polarization ellipses of the four secondary branches is $\Delta\phi=\Delta\omega(t_1-t_0)$ as the waves pass through the time wave plate (see Fig.~\ref{fig1}).  And in terms of the boundary conditions at $t_1$, the relative lengths of the two polarized axes satisfy
\begin{align}
  a^{\sigma\tau}=&\delta^\sigma\left(\frac{\delta^\tau\omega_+\omega_--{\omega_0}^2}{2\omega_0\omega_+\omega_-}\bar\omega+\frac{1-\delta^\tau}{2}\right),\\
  b^{\sigma\tau}=&-\delta^\sigma\frac{\delta^\tau\omega_+\omega_-+{\omega_0}^2}{2\omega_0\omega_+\omega_-}\Delta\omega.
\end{align}

\vspace{15pt}
\noindent\textbf{\large Velocities of TCB modes.\ }
The phase velocities and the group velocities of two TCB modes are, respectively,
\begin{align}\label{velocity}
   v_{\rm p\pm}=&\frac{\omega_\pm}{k}=v_1\sqrt{\frac{k\pm\beta}{k}},\\
   \bm{v}_{\rm g\pm}=&\nabla_{\bm{k}}\,\omega_{\pm}=v_1\frac{k\pm\beta/2}{\sqrt{k(k\pm\beta)}}\bm{\hat{k}}.
\end{align}
As noted in Ref.~\cite{carroll1990limits,itin2004carroll,itin2008wave}, the two phase velocities meet $v_{\rm p+}>v_1>v_{\rm p-}$, and the two group velocities meet $v_{\rm g\pm}>v_1$. For the axion field in vacuum, $v_{\rm p+}$ and $v_{\rm g\pm}$ always exceed the speed of light $c$ in vacuum. Though $v_1<c$ in real media, $v_{\rm p+}$ and $v_{\rm g\pm}$ will be still superluminal when $k\rightarrow0$ for $v_{\rm p+},\ v_{\rm g+}$ and $k\rightarrow\beta$ for $v_{\rm g-}$. However, neither phase velocity nor group velocity represents the true velocity of energy or information transfer, therefore the superluminal effects of these two types of velocities do not violate the causality and have been observed in various experiments \cite{wang2000gain,alexeev2002measurement,brunner2004direct}. By means of the average Poynting vector and energy density over a period,  we also can calculate the energy transport velocities of the two TCB states
\begin{equation}\label{energy velocity}
  \bm{v}_{\rm E\pm}=\frac{\langle\bm{S}_\pm\rangle}{\langle W_\pm\rangle}=v_1\frac{\sqrt{k(k\pm\beta)}}{k\pm\beta/2}\bm{\hat{k}}.
\end{equation}
On the contrary to the group velocities, $v_{\rm E\pm}$ are always less than $v_1$.
Moreover, we prove that the front velocity $v_{\rm f}$ (the velocity of wave front which represents the speed of information propagation) of the two TCB modes is precisely $v_1$, when only concerning the dispersion caused by the constant rate $\beta$ of the ME coefficients but regardless of the dispersion of $\varepsilon,\ \mu,\ \beta$ with respect to wave number $k$ (the detailed derivation is given in the supplementary information). 
Therefore, neither energy nor information of TCB modes propagates superluminally. The comparison of four types of velocities is shown in Fig.~\ref{fig_velocity}\,(b).

\vspace{15pt}
\noindent\textbf{\large Gaussian pulse traversing a time interface.\ }
The plane wave solutions we have discussed are widespread in the whole space. However, the time wave plate made of time dependent media should only have a finite scale in practice. 
We accordingly need to analyze the propagation of wave packages with finite length. Consider a Gaussian pulse $  \bm{B}^{\rm in}_\pm=A^0_\pm\exp\left[-(z-v_0t)^2/a^2+\mi k_0(z-v_0t)\right]\bm{\hat{U}}_\mp$
with left or right circular polarization and width $a$ incident onto the time interface $t_0$ of a time wave plate. Here, we still only concern the dispersion caused by $\beta$. Taking account of the temporal boundary conditions, we obtain the magnetic fields, for $t>t_0$,
\begin{equation}\label{pulse in time plate}
  \bm{B}_\pm=A^0_\pm\left(B_\pm^1+B_\pm^2+B_\pm^3\right)\bm{\hat{U}}_\mp,
\end{equation}
where $B_\pm^1,\ B_\pm^2$ denote the time refraction and reflection parts respectively, and $B_\pm^3$ denotes the non-traveling wave part. The three parts of $\bm{B}_-$ take the forms
\begin{align}
\begin{split}\label{pulse refraction and reflection}
  B_-^\sigma =&\frac{a }{2\sqrt{\pi}}\left(\int_{-\infty}^0+\int_\beta^{\infty}\right)\dif k \frac{\omega_-(k)-\delta^\sigma v_0k}{2\omega_-(k)}
  \exp\left[-\frac{a^2(k-k_0)^2}{4}+\mi\Big(kz'+\delta^\sigma \omega_-(k)t'\Big)\right],
  \quad(\sigma=1,2),
\end{split}\\
\begin{split}
  B_-^3=&\frac{a }{2\sqrt{\pi}}\int_0^\beta\dif k \sum_{\sigma=1}^2\frac{\tilde\omega_-(k)+\mi\delta^\sigma v_0k}{2\tilde\omega_-(k)}
  \exp\left[-\frac{a^2(k-k_0)^2}{4}+\mi kz'-\delta^\sigma \tilde\omega_-(k)t'\right],
\end{split}
\end{align}
with $\tilde\omega_-(k)=v_0\sqrt{k(\beta-k)}$ and $z'=z-v_0t_0$.  And the three parts of $\bm{B}_+$ have  similar expressions.

\begin{figure*}[t]
  \centering
\includegraphics[width=1\columnwidth,clip]{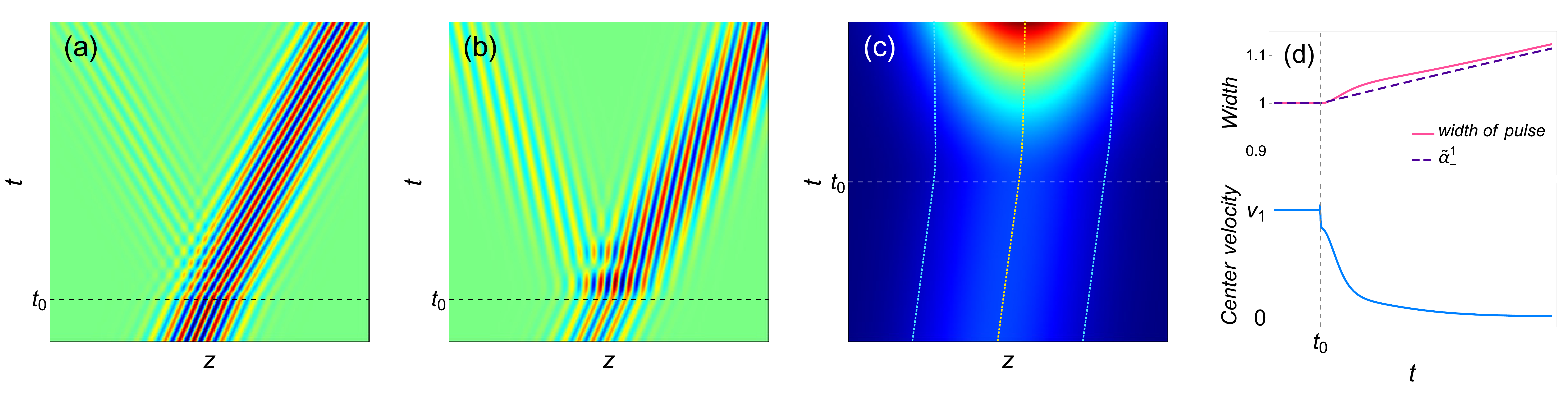}
\caption{\label{fig_pulse}Magnetic field patterns of circularly polarized states (a) $\bm{B}_+$ and (b) $\bm{B}_-$ in spacetime for a corresponding polarized Gaussian pulse incident onto the time interface $t_0$, whose $k$ spectrum mainly concentrates in the traveling-wave band, namely $k_0-\beta\gg2/a$. (c) $|\bm{B}_-|$ of a pulse whose $k$ spectrum mainly concentrates in the non-traveling- wave band, namely $2/a\ll k_0\ll\beta-2/a$. In this case, the pulse is nearly trapped while its intensity increases rapidly. The yellow dashed curve and the two light-blue dashed curves trace the center of the pulse and the edges of the pulse respectively. (d) The width (scale of 1 to the original width $2a$)  and the center velocity of the pulse varying with time.}
\end{figure*}

For the situation $k_0-\beta\gg2/a$, the non-traveling wave part $B_\pm^3$ can be neglected, and the range of integration in Eq.~(\ref{pulse refraction and reflection}) can approximate to $-\infty$ to $\infty$. In addition, we expand $\omega_\pm(k)$ near the center wave number $k_0$ in a Taylor series
 \begin{equation}
 \begin{split}
 \omega_\pm(k_0+\kappa)
 =&\omega_\pm(k_0)+\left.\partial_k\omega_\pm\right|_{k_0}\kappa+\frac{1}{2}\left.\partial^2_k\omega_\pm\right|_{k_0}\kappa^2+\cdots\\
 =&\omega_\pm(k_0)+v_{\rm g\pm}(k_0)\kappa+\frac{1}{2}\left.\partial_kv_{\rm g\pm}\right|_{k_0}\kappa^2+\cdots,
\end{split}
\end{equation}
 and neglect the high order terms (order$\geq 3$), then the refracted and reflected pulses have the approximations:
\begin{align}\label{pulse refraction and reflection 2}
  B_\pm^\sigma=&\frac{a}{2\alpha^\sigma_\pm}\left[1-\delta^\sigma\sqrt{\frac{k_0}{k_0\pm\beta}}\bigg(1\pm \frac{\mi\,\beta}{k_0(k_0\pm\beta)}\frac{{\zeta^\sigma_\pm}}{{\alpha^\sigma_\pm}^2}\bigg)\right]
  \exp\left[-\frac{{\zeta^\sigma_\pm}^2}{{\alpha^\sigma_\pm}^2}+\mi\Big(k_0z'+\delta^\sigma\omega_\pm(k_0)t'\Big)\right],
\end{align}
where $\zeta^\sigma_\pm=z'+\delta^\sigma v_{\rm g\pm}(k_0)t'$ is the relative coordinate with respect to the center of the wave package, and $\alpha^\sigma_\pm=\sqrt{a^2-2\mi\,\delta^\sigma \partial_k v_{\rm g\pm}|_{k_0} t'}$. The time refractions and time reflections for two different circularly polarized pulses are shown in Fig.~\ref{fig_pulse}(a) and Fig.~\ref{fig_pulse}(b).
Actually, this approximation is valid only when $t\ll a^2(k_0-\beta)^2/(4v_0\beta)$, because $B^3_\pm$ increases exponentially. However, as $k_0-\beta\gg2/a$, the upper bound of time could be a long period.
According to Eq.~(\ref{pulse refraction and reflection 2}), the term proportional to $\mi\beta$ is extremely small in the main range of the pulses $|z-\zeta^\sigma_\pm|\sim a$. Omitting this term, it is clear that the pulse propagates with group velocity $v_{\rm g\pm}$, and the dispersion of $v_{\rm g\pm}$ induces the pulse width to change with time.

For another particular case $2/a\ll k_0\ll\beta-2/a$, the traveling parts of refraction and reflection $B^1_+,\ B^2_+$ shown in Eq.~(\ref{pulse refraction and reflection 2}) still offer the major contribution to $\bm{B}_+$. However, $\bm{B}_-$ mainly concentrates in the non-traveling part,  ignoring the refraction and reflection parts $B^1_-,\ B^2_-$ is thus reasonable, and the approximate solution reads
\begin{align}
 B_-^3=&\sum_{\sigma=1}^2\frac{a}{2\tilde\alpha^\sigma_-}\left[1+\mi\,\delta^\sigma\sqrt{\frac{k_0}{\beta-k_0}}\bigg(1+ \frac{\beta}{k_0(\beta-k_0)}\frac{{\tilde\zeta^\sigma_-}}{{{}\tilde\alpha^\sigma_-}^2}\bigg)\right]  \exp\left[\frac{{{}\tilde\zeta^\sigma_-}^2}{{{}\tilde\alpha^\sigma_-}^2}+\mi k_0z'-\delta^\sigma\tilde\omega_-(k_0)t'\right],
\end{align}
with $\tilde\zeta^\sigma_-=\mi z'-\delta^\sigma \partial_k\tilde\omega_- |_{k_0}t'$,  $\tilde\alpha^\sigma_-=\sqrt{a^2+2\delta^\sigma \partial^2_k\tilde\omega_-|_{k_0} t'}$. Fig.~\ref{fig_pulse}(c) shows the pattern of $|\bm{B}_-|$ as the pulse traversing the time interface. Fig.~\ref{fig_pulse}(d) plots the velocity of the pulse center and the width of the pulse (defined as the distance between the two edges where $|\bm{B}_-|$ equal to $1/\me$ times $|\bm{B}_-|$ at the center of the pulse) changing with time. Consequently, the pulse keeps nearly fixed center after traversing the time interface, while its intensity increases with the magnitude about $\exp\big(t^2)$. The width of the pulse increases with time, and it can be characterized by $\tilde{\alpha}^1_-$ approximately as shown in Fig.~\ref{fig_pulse}(d).

\begin{figure}[t]
\includegraphics[width=0.65\columnwidth,clip]{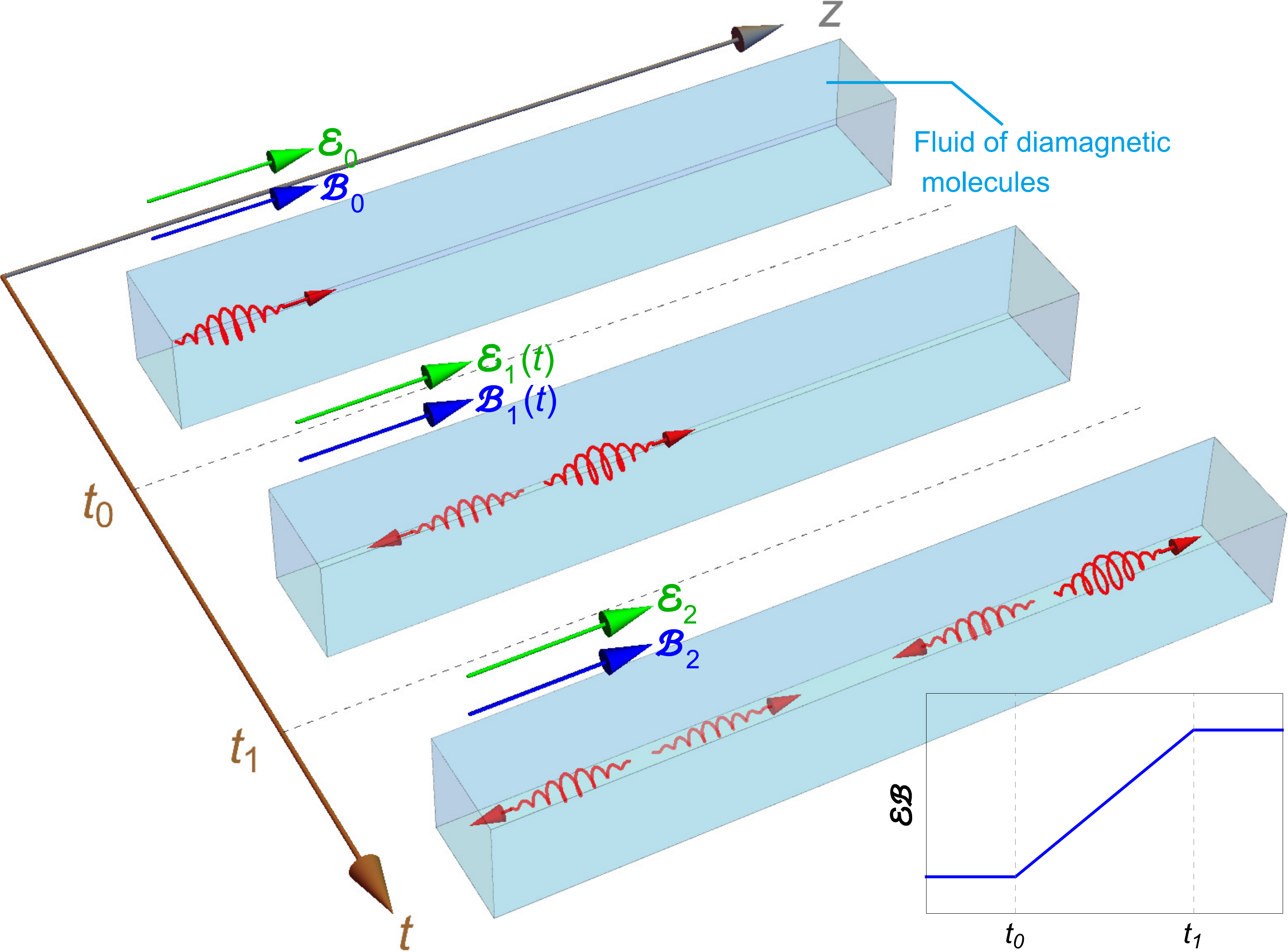}
\caption{\label{exp}Illustration of circularly polarized pulse traveling in a fluid of diamagnetic molecules located in time-dependent external electric field $\mathcal{E}$ and magnetic field $\mathcal{B}$ both parallel to the propagating direction of the pulse. For the three-piece product $\mathcal{E\,B}$, the fluid acts as a time wave plate. At $t_0$, the incident pulse splits into a refracted one and a reflected one. At $t_1$, the two pulses further split into four.}
\end{figure}

\vspace{15pt}
\noindent\textbf{\large Experimental design.\ }
Considering a fluid in the presence of external electric and magnetic fields, the multipolar polarizations induced by external electric or magnetic fields can cause the fluid to be anisotropic and lead to Kerr effect or Cotton-Mouton effect. More specially, a parallel pair of external electric field $\bm{\mathcal{E}}$ and magnetic field $\bm{\mathcal{B}}$ will induce the Jones birefringence for a light beam propagating perpendicularly to the direction of the fields \cite{graham1983jones,roth2000observation,roth2000Magneto-electric,rizzo2003jones}. The Jones birefringence has been shown to be a bianisotropic effect \cite{roth2000Magneto-electric}. For symmetric analysis, the external electric field $\bm{\mathcal{E}}$ is $P$ odd, and the external magnetic field $\bm{\mathcal{B}}$ is $T$ odd, but the parallelism of the two fields protects the combined $PT$ symmetry. This fact indicates the existence of the axion type ME coupling as we have mentioned. The ME coupling tensor of molecules can be expanded with respect to the external fields
\begin{equation}
\begin{split}
  G^i_j(\bm{\mathcal{E}},\bm{\mathcal{B}})
  =G^i_j+G^i_{jk}\mathcal{B}^k+G^{ik}_{jl}\mathcal{E}_k\mathcal{B}^l+\cdots.
\end{split}
\end{equation}
The coefficients of each order are determined by solving the time-dependent perturbation of the molecular hamiltonian \cite{rizzo2003jones}. The Boltzmann average over all orientations of diamagnetic molecules yields \cite{graham1983jones,rizzo2003jones}
\begin{equation}
  {\xi^i}_j={\zeta_j}^i=\langle G^i_j(\bm{\mathcal{E}},\bm{\mathcal{B}}) \rangle =N\mathcal{E}_z\mathcal{B}_z\Big\langle G^{iz}_{jz}+\frac{\mu_{\rm e}^z}{k_{\rm B}T}G^{i}_{jz}\Big\rangle,
\end{equation}
where the external fields are supposed to parallel $z$ axis, $N$ is the number density of molecules, $\mu_{\rm e}^z$ is the $z$ component of the permanent molecular electric dipole moment, $k_{\rm B}$ is the Boltzmann constant, and $T$ is temperature. Since the system is symmetric with respect to $z$ axis, the medium should retain isotropic in the $x-y$ plane and has a uniaxial ME tensor $\matr\xi=\matr\zeta=\mathrm{diag}(\Theta_\perp,\Theta_\perp,\Theta_\parallel)$.  Thus a beam propagating perpendicularly to $z$ axis has two Jones birefringent eigenmodes, linearly polarized along the $\pm\pi/4$ directions with respect to $z$ axis respectively, with the difference of refractive indexes $\Delta n_{\rm J}=c\mu(\Theta_\parallel-\Theta_\perp)=c\mu(\alpha_\parallel-\alpha_\perp)\mathcal{E}_z\mathcal{B}_z=c\mu\alpha_{\rm J}\mathcal{E}_z\mathcal{B}_z$~\cite{roth2000Magneto-electric}. 
However, if a transverse polarized light travels along $z$ axis, {\it i.e.} parallel to the external fields, it will experience the isotropic axion-type ME coupling $\Theta_\perp=\alpha_\perp\mathcal{E}_z\mathcal{B}_z$. In terms of isotropic average~ \cite{graham1983jones}, the ME coefficient in $x-y$ plane, is
\begin{equation}
\begin{split}
  \Theta_\perp  &=\frac{N\mathcal{E}_z\mathcal{B}_z}{30}\left[ 4G^{ij}_{ij}-G^{ij}_{ji}-g_{ij}g^{kl}G^{ij}_{kl}+\frac{\mu_{\rm e}^i}{k_{\rm B}T}\left(4G^{j}_{ji}-G^{j}_{ij}-g_{ij}g^{kl}G^{j}_{kl}\right)\right].
\end{split}
\end{equation}
As a result, the effective axion field can be controlled via the external electric and magnetic fields.  
If the product of the external fields $\mathcal{E}_z\mathcal{B}_z$ changes with time, we could observe the TCB and correlated phenomena predicted in this paper. The schematic illustration are shown in Fig.~\ref{exp}.

In principle, the TCB, as well as the ME coupling, caused by the time-varying external fields can arise in all media, while its magnitude is characterized by $\beta=\mu\dot{\Theta}_\perp=\mu\alpha_\perp\dif(\mathcal{E}_z\mathcal{B}_z)/\dif t$. Supposing the product of the fields varies linearly with time, the magnitude is determined by two parts, one is the intrinsic property of the medium $\alpha_\perp$, the other is the rate of  field change $\Delta(\mathcal{E}_z\mathcal{B}_z)/\Delta t$. In the first order approximation, the frequencies and the phase velocities of the two TCB modes are $\omega_\pm=\omega_0[1\pm\beta/(2k)]$ and $v_{\rm p\pm}=v_1[1\pm\beta/(2k)]$ respectively. And the refractive-index difference of the two TCB modes is 
\begin{equation}
\Delta n_\pm=n_+-n_-=\frac{c\beta}{\omega_0}=\frac{c\mu\alpha_\perp\mathcal{E}_z\mathcal{B}_z}{\omega_0\Delta t}\sim\frac{\Delta n_{\rm J}}{\omega_0\Delta t},
\end{equation}
with the assumption that the product of the external fields increases linearly from $0$ to the the final value $\mathcal{E}_z\mathcal{B}_z$ in the time interval $\Delta t$. Here, the symbol ``$\sim$'' means the quantities of two sides have the same order of magnitude, since $\alpha_\perp$ and $\alpha_\parallel$ are generically in the same order. 

According to the experimental results in Ref.~\cite{roth2000observation,roth2000Magneto-electric}, molecules with a low-lying strong charge transfer transition of approximately octupolar symmetry and a permanent electric dipole moment will have relative large ME coupling. In this experiment, the Jones birefringence are observed in three typical molecular liquids, namely methylcyclopentadienyl-Mn-tricarbonyl, cyclohexadienyl-Fe-tricarbonyl, and Ti-bis(ethyl-acetoacetato) diisopropoxide, with the magnitude about $\Delta n_{\rm J}\sim10^{-11}$ under the parameters $\omega_0=2.979\times 10^3\mathrm{THz}$ (HeNe laser), $\mathcal{E}_z\sim 2\times10^5\mathrm{V/m}$, $\mathcal{B}_z\sim15\mathrm{T}$ at room temperature and $1\mathrm{atm}$. Adopting these experimental parameters and assuming the time interval of field change $\Delta t\sim 10^{-9}\mathrm{s}$ (the characteristic frequency of the external fields is equivalent to $\mathrm{GHz}$),  we can estimate the refractive-index difference of the two TCB modes $\Delta n_\pm\sim 10^{-17}$. On the other hand, previous experiments for small birefringence measurements have achieved the sensitivity $\Delta n\sim 10^{-18}$ via the metrology of high finesse resonant cavity~\cite{bailly2010highly,robilliard2010towards,pelle2011magnetoelectric}, to measure the TCB effect is accordingly feasible. 
Since the group velocities of the two TCB modes are nearly equal $v_{\rm g\pm}=v_1[1+\beta^2/(8k^2)]$ for small $\beta$, we can ignore the central separation of two superposed TCB pulses during the time interval $\Delta t$ and regard them as a single pulse with the TFR $\Delta\phi=\Delta\omega\Delta t\sim 10^{-11}\mathrm{rad}$ which is large enough for detection  as a $10^{-13}\mathrm{rad}$ resolution of phase shift  has been achieved experimentally~\cite{durand2010shot}.

If the external fields are both parallel to the propagating direction of the pulse rigorously, no other birefringent effects that can disturb the observation of TCB, {\it e.g.} Kerr or Cotton-Mouton effects,  would arise. However, the time dependence of the external fields will induce fields in the $x-y$ plane inevitably. Supposing only $\mathcal{E}_z$ changes with time but $\mathcal{B}_z$ is constant, the linearly varying $\mathcal{E}_z$ induces an eddy magnetic field $\mathcal{B}_\theta=(\mathcal{E}_z/\Delta t)r/(2c^2)$ around $z$ axis, and $\mathcal{B}_\theta\sim 10^{-5}\mathrm{T}$ in the area of $r<10^{-2}\mathrm{m}$ which is thus small enough to be ignored. For experimental setup, a big challenge is to precisely control the external fields. Theoretically, the external fields at any locations should change simultaneously in the laboratory reference system, namely the variation of $\mathcal{E}_z\mathcal{B}_z$ at different points is spacelike, since the effective axion field $\Theta_\perp$ only depends on time. In practice, the speed of light in the media is less than vacuum, thus the prerequisite could be relaxed into that the fields begin to change before the pulse arrives. If there is a slow-light system with strong ME coupling $\alpha_\perp$, then the technical requirement could be largely reduced.

\vspace{20pt}
\noindent
\textbf{\Large Conclusion}\\
\noindent
To summarize, we demonstrate that light with a certain wave vector $\bm{k}$ always corresponds to a pair of circularly polarized modes, {\it i.e.} the TCB modes, in time-dependent axion-type ME media. We study the time refraction and time reflection of plane waves and Gaussian pulses traveling in this type of media, and predict the time Faraday effect as a consequence of the superposition of the two TCB modes. We also propose a scheme to realize TCB in practice. According to our estimations with the realistic parameters, the magnitude of TCB is observable via existing experimental techniques. As the significance but difficulty for detecting axion particles, our proposal offers an alternative way to simulate and study the interaction of light with time-dependent axion field. On the other hand, by exchanging the status of space and time, we foresee that various effects in space-dependent media would have their temporal counterparts in time-dependent media for not only electromagnetic fields but also all kinds of waves. We hope our work could inspire more research in this novel area.



\begin{thebibliography}{0}%
\makeatletter
\providecommand \@ifxundefined [1]{%
 \@ifx{#1\undefined}
}%
\providecommand \@ifnum [1]{%
 \ifnum #1\expandafter \@firstoftwo
 \else \expandafter \@secondoftwo
 \fi
}%
\providecommand \@ifx [1]{%
 \ifx #1\expandafter \@firstoftwo
 \else \expandafter \@secondoftwo
 \fi
}%
\providecommand \natexlab [1]{#1}%
\providecommand \enquote  [1]{``#1''}%
\providecommand \bibnamefont  [1]{#1}%
\providecommand \bibfnamefont [1]{#1}%
\providecommand \citenamefont [1]{#1}%
\providecommand \href@noop [0]{\@secondoftwo}%
\providecommand \href [0]{\begingroup \@sanitize@url \@href}%
\providecommand \@href[1]{\@@startlink{#1}\@@href}%
\providecommand \@@href[1]{\endgroup#1\@@endlink}%
\providecommand \@sanitize@url [0]{\catcode `\\12\catcode `\$12\catcode
  `\&12\catcode `\#12\catcode `\^12\catcode `\_12\catcode `\%12\relax}%
\providecommand \@@startlink[1]{}%
\providecommand \@@endlink[0]{}%
\providecommand \url  [0]{\begingroup\@sanitize@url \@url }%
\providecommand \@url [1]{\endgroup\@href {#1}{\urlprefix }}%
\providecommand \urlprefix  [0]{URL }%
\providecommand \Eprint [0]{\href }%
\providecommand \doibase [0]{http://dx.doi.org/}%
\providecommand \selectlanguage [0]{\@gobble}%
\providecommand \bibinfo  [0]{\@secondoftwo}%
\providecommand \bibfield  [0]{\@secondoftwo}%
\providecommand \translation [1]{[#1]}%
\providecommand \BibitemOpen [0]{}%
\providecommand \bibitemStop [0]{}%
\providecommand \bibitemNoStop [0]{.\EOS\space}%
\providecommand \EOS [0]{\spacefactor3000\relax}%
\providecommand \BibitemShut  [1]{\csname bibitem#1\endcsname}%
\let\auto@bib@innerbib\@empty
\end{thebibliography}%


\begin{thebibliography}{10}
\expandafter\ifx\csname url\endcsname\relax
  \def\url#1{\texttt{#1}}\fi
\expandafter\ifx\csname urlprefix\endcsname\relax\def\urlprefix{URL }\fi
\providecommand{\bibinfo}[2]{#2}
\providecommand{\eprint}[2][]{\url{#2}}

\bibitem{dzyaloshinskii1960magneto}
\bibinfo{author}{Dzyaloshinskii, I.}
\newblock \bibinfo{title}{On the magneto-electrical effect in
  antiferromagnets}.
\newblock \emph{\bibinfo{journal}{Sov. Phys. JETP}}
  \textbf{\bibinfo{volume}{10}}, \bibinfo{pages}{628--629}
  (\bibinfo{year}{1960}).

\bibitem{folen1961anisotropy}
\bibinfo{author}{Folen, V.~J.}, \bibinfo{author}{Rado, G.~T.} \&
  \bibinfo{author}{Stalder, E.~W.}
\newblock \bibinfo{title}{Anisotropy of the magnetoelectric effect in $\rm
  \uppercase{C}r_2 \uppercase{O}_3$}.
\newblock \emph{\bibinfo{journal}{Phys. Rev. Lett.}}
  \textbf{\bibinfo{volume}{6}}, \bibinfo{pages}{607} (\bibinfo{year}{1961}).

\bibitem{landau1984electrodynamics}
\bibinfo{author}{Landau, L.~D.}, \bibinfo{author}{Lifshitz, E.~M.} \&
  \bibinfo{author}{Pitaevskii, L.~P.}
\newblock \emph{\bibinfo{title}{Electrodynamics of continuous media}}
  (\bibinfo{publisher}{Elsevier}, \bibinfo{year}{1984}).

\bibitem{fiebig2005revival}
\bibinfo{author}{Fiebig, M.}
\newblock \bibinfo{title}{Revival of the magnetoelectric effect}.
\newblock \emph{\bibinfo{journal}{J. Phys. D: Appl. Phys.}}
  \textbf{\bibinfo{volume}{38}}, \bibinfo{pages}{R123} (\bibinfo{year}{2005}).

\bibitem{eerenstein2006multiferroic}
\bibinfo{author}{Eerenstein, W.}, \bibinfo{author}{Mathur, N.~D.} \&
  \bibinfo{author}{Scott, J.~F.}
\newblock \bibinfo{title}{Multiferroic and magnetoelectric materials}.
\newblock \emph{\bibinfo{journal}{Nature}} \textbf{\bibinfo{volume}{442}},
  \bibinfo{pages}{759--765} (\bibinfo{year}{2006}).

\bibitem{pyatakov2012magnetoelectric}
\bibinfo{author}{Pyatakov, A.~P.} \& \bibinfo{author}{Zvezdin, A.~K.}
\newblock \bibinfo{title}{Magnetoelectric and multiferroic media}.
\newblock \emph{\bibinfo{journal}{Phys. Usp.}} \textbf{\bibinfo{volume}{55}},
  \bibinfo{pages}{557--581} (\bibinfo{year}{2012}).

\bibitem{qi2008topological}
\bibinfo{author}{Qi, X.-L.}, \bibinfo{author}{Hughes, T.~L.} \&
  \bibinfo{author}{Zhang, S.-C.}
\newblock \bibinfo{title}{Topological field theory of time-reversal invariant
  insulators}.
\newblock \emph{\bibinfo{journal}{Phys. Rev. B}} \textbf{\bibinfo{volume}{78}},
  \bibinfo{pages}{195424} (\bibinfo{year}{2008}).

\bibitem{li2010dynamical}
\bibinfo{author}{Li, R.}, \bibinfo{author}{Wang, J.}, \bibinfo{author}{Qi,
  X.-L.} \& \bibinfo{author}{Zhang, S.-C.}
\newblock \bibinfo{title}{Dynamical axion field in topological magnetic
  insulators}.
\newblock \emph{\bibinfo{journal}{Nat. Phys.}} \textbf{\bibinfo{volume}{6}},
  \bibinfo{pages}{284--288} (\bibinfo{year}{2010}).

\bibitem{qi2009inducing}
\bibinfo{author}{Qi, X.-L.}, \bibinfo{author}{Li, R.}, \bibinfo{author}{Zang,
  J.} \& \bibinfo{author}{Zhang, S.-C.}
\newblock \bibinfo{title}{Inducing a magnetic monopole with topological surface
  states}.
\newblock \emph{\bibinfo{journal}{Science}} \textbf{\bibinfo{volume}{323}},
  \bibinfo{pages}{1184--1187} (\bibinfo{year}{2009}).

\bibitem{essin2009magnetoelectric}
\bibinfo{author}{Essin, A.~M.}, \bibinfo{author}{Moore, J.~E.} \&
  \bibinfo{author}{Vanderbilt, D.}
\newblock \bibinfo{title}{Magnetoelectric polarizability and axion
  electrodynamics in crystalline insulators}.
\newblock \emph{\bibinfo{journal}{Phys. Rev. Lett.}}
  \textbf{\bibinfo{volume}{102}}, \bibinfo{pages}{146805}
  (\bibinfo{year}{2009}).

\bibitem{nomura2011surface}
\bibinfo{author}{Nomura, K.} \& \bibinfo{author}{Nagaosa, N.}
\newblock \bibinfo{title}{Surface-quantized anomalous hall current and the
  magnetoelectric effect in magnetically disordered topological insulators}.
\newblock \emph{\bibinfo{journal}{Phys. Rev. Lett.}}
  \textbf{\bibinfo{volume}{106}}, \bibinfo{pages}{166802}
  (\bibinfo{year}{2011}).

\bibitem{de2001post}
\bibinfo{author}{de~Lange, O.~L.} \& \bibinfo{author}{Raab, R.~E.}
\newblock \bibinfo{title}{Post's constraint for electromagnetic constitutive
  relations}.
\newblock \emph{\bibinfo{journal}{J. Opt. A: Pure Appl. Opt.}}
  \textbf{\bibinfo{volume}{3}}, \bibinfo{pages}{L23} (\bibinfo{year}{2001}).

\bibitem{hehl2005linear}
\bibinfo{author}{Hehl, F.~W.} \& \bibinfo{author}{Obukhov, Y.~N.}
\newblock \bibinfo{title}{Linear media in classical electrodynamics and the
  post constraint}.
\newblock \emph{\bibinfo{journal}{Phys. Lett. A}}
  \textbf{\bibinfo{volume}{334}}, \bibinfo{pages}{249--259}
  (\bibinfo{year}{2005}).

\bibitem{obukhov2005measuring}
\bibinfo{author}{Obukhov, Y.~N.} \& \bibinfo{author}{Hehl, F.~W.}
\newblock \bibinfo{title}{Measuring a piecewise constant axion field in
  classical electrodynamics}.
\newblock \emph{\bibinfo{journal}{Phys. Lett. A}}
  \textbf{\bibinfo{volume}{341}}, \bibinfo{pages}{357--365}
  (\bibinfo{year}{2005}).

\bibitem{jung2004optical}
\bibinfo{author}{Jung, J.~H.} \emph{et~al.}
\newblock \bibinfo{title}{Optical magnetoelectric effect in the polar $\rm
  \uppercase{G}a\uppercase{F}e\uppercase{O}_3$ ferrimagnet}.
\newblock \emph{\bibinfo{journal}{Phys. Rev. Lett.}}
  \textbf{\bibinfo{volume}{93}}, \bibinfo{pages}{037403}
  (\bibinfo{year}{2004}).

\bibitem{kida2006enhanced}
\bibinfo{author}{Kida, N.} \emph{et~al.}
\newblock \bibinfo{title}{Enhanced optical magnetoelectric effect in a
  patterned polar ferrimagnet}.
\newblock \emph{\bibinfo{journal}{Phys. Rev. Lett.}}
  \textbf{\bibinfo{volume}{96}}, \bibinfo{pages}{167202}
  (\bibinfo{year}{2006}).

\bibitem{kida2007optical}
\bibinfo{author}{Kida, N.} \emph{et~al.}
\newblock \bibinfo{title}{Optical magnetoelectric effect of patterned oxide
  superlattices with ferromagnetic interfaces}.
\newblock \emph{\bibinfo{journal}{Phys. Rev. Lett.}}
  \textbf{\bibinfo{volume}{99}}, \bibinfo{pages}{197404}
  (\bibinfo{year}{2007}).

\bibitem{takahashi2012magnetoelectric}
\bibinfo{author}{Takahashi, Y.}, \bibinfo{author}{Shimano, R.},
  \bibinfo{author}{Kaneko, Y.}, \bibinfo{author}{Murakawa, H.} \&
  \bibinfo{author}{Tokura, Y.}
\newblock \bibinfo{title}{Magnetoelectric resonance with electromagnons in a
  perovskite helimagnet}.
\newblock \emph{\bibinfo{journal}{Nat. Phys.}} \textbf{\bibinfo{volume}{8}},
  \bibinfo{pages}{121--125} (\bibinfo{year}{2012}).

\bibitem{kamenetskii2009tellegen}
\bibinfo{author}{Kamenetskii, E.}, \bibinfo{author}{Sigalov, M.} \&
  \bibinfo{author}{Shavit, R.}
\newblock \bibinfo{title}{Tellegen particles and magnetoelectric
  metamaterials}.
\newblock \emph{\bibinfo{journal}{J. Appl. Phys.}}
  \textbf{\bibinfo{volume}{105}}, \bibinfo{pages}{013537}
  (\bibinfo{year}{2009}).

\bibitem{tse2010giant}
\bibinfo{author}{Tse, W.-K.} \& \bibinfo{author}{MacDonald, A.~H.}
\newblock \bibinfo{title}{Giant magneto-optical kerr effect and universal
  faraday effect in thin-film topological insulators}.
\newblock \emph{\bibinfo{journal}{Phys. Rev. Lett.}}
  \textbf{\bibinfo{volume}{105}}, \bibinfo{pages}{057401}
  (\bibinfo{year}{2010}).

\bibitem{tse2010magneto}
\bibinfo{author}{Tse, W.-K.} \& \bibinfo{author}{MacDonald, A.~H.}
\newblock \bibinfo{title}{Magneto-optical and magnetoelectric effects of
  topological insulators in quantizing magnetic fields}.
\newblock \emph{\bibinfo{journal}{Phys. Rev. B}} \textbf{\bibinfo{volume}{82}},
  \bibinfo{pages}{161104} (\bibinfo{year}{2010}).

\bibitem{tse2011magneto}
\bibinfo{author}{Tse, W.-K.} \& \bibinfo{author}{MacDonald, A.~H.}
\newblock \bibinfo{title}{Magneto-optical \uppercase{F}araday and
  \uppercase{K}err effects in topological insulator films and in other layered
  quantized hall systems}.
\newblock \emph{\bibinfo{journal}{Phys. Rev. B}} \textbf{\bibinfo{volume}{84}},
  \bibinfo{pages}{205327} (\bibinfo{year}{2011}).

\bibitem{Bliokh2014magnetoelectric}
\bibinfo{author}{Bliokh, K.~Y.}, \bibinfo{author}{Kivshar, Y.~S.} \&
  \bibinfo{author}{Nori, F.}
\newblock \bibinfo{title}{Magnetoelectric effects in local light-matter
  interactions}.
\newblock \emph{\bibinfo{journal}{Phys. Rev. Lett.}}
  \textbf{\bibinfo{volume}{113}}, \bibinfo{pages}{033601}
  (\bibinfo{year}{2014}).

\bibitem{wilczek1987two}
\bibinfo{author}{Wilczek, F.}
\newblock \bibinfo{title}{Two applications of axion electrodynamics}.
\newblock \emph{\bibinfo{journal}{Phys. Rev. Lett.}}
  \textbf{\bibinfo{volume}{58}}, \bibinfo{pages}{1799} (\bibinfo{year}{1987}).

\bibitem{peccei1977cp}
\bibinfo{author}{Peccei, R.~D.} \& \bibinfo{author}{Quinn, H.~R.}
\newblock \bibinfo{title}{\uppercase{CP} conservation in the presence of
  pseudoparticles}.
\newblock \emph{\bibinfo{journal}{Phys. Rev. Lett.}}
  \textbf{\bibinfo{volume}{38}}, \bibinfo{pages}{1440} (\bibinfo{year}{1977}).

\bibitem{carroll1990limits}
\bibinfo{author}{Carroll, S.~M.}, \bibinfo{author}{Field, G.~B.} \&
  \bibinfo{author}{Jackiw, R.}
\newblock \bibinfo{title}{Limits on a lorentz-and parity-violating modification
  of electrodynamics}.
\newblock \emph{\bibinfo{journal}{Phys. Rev. D}} \textbf{\bibinfo{volume}{41}},
  \bibinfo{pages}{1231} (\bibinfo{year}{1990}).

\bibitem{itin2004carroll}
\bibinfo{author}{Itin, Y.}
\newblock \bibinfo{title}{Carroll-\uppercase{F}ield-\uppercase{J}ackiw
  electrodynamics in the premetric framework}.
\newblock \emph{\bibinfo{journal}{Phys. Rev. D}} \textbf{\bibinfo{volume}{70}},
  \bibinfo{pages}{025012} (\bibinfo{year}{2004}).

\bibitem{itin2008wave}
\bibinfo{author}{Itin, Y.}
\newblock \bibinfo{title}{Wave propagation in axion electrodynamics}.
\newblock \emph{\bibinfo{journal}{Gen. Relativ. Gravit.}}
  \textbf{\bibinfo{volume}{40}}, \bibinfo{pages}{1219--1238}
  (\bibinfo{year}{2008}).

\bibitem{ginis2010frequency}
\bibinfo{author}{Ginis, V.}, \bibinfo{author}{Tassin, P.},
  \bibinfo{author}{Craps, B.} \& \bibinfo{author}{Veretennicoff, I.}
\newblock \bibinfo{title}{Frequency converter implementing an optical analogue
  of the cosmological redshift}.
\newblock \emph{\bibinfo{journal}{Opt. Express}} \textbf{\bibinfo{volume}{18}},
  \bibinfo{pages}{5350--5355} (\bibinfo{year}{2010}).

\bibitem{cummer2011frequency}
\bibinfo{author}{Cummer, S.~A.} \& \bibinfo{author}{Thompson, R.~T.}
\newblock \bibinfo{title}{Frequency conversion by exploiting time in
  transformation optics}.
\newblock \emph{\bibinfo{journal}{J. Opt.}} \textbf{\bibinfo{volume}{13}},
  \bibinfo{pages}{024007} (\bibinfo{year}{2011}).

\bibitem{mccall2011spacetime}
\bibinfo{author}{McCall, M.~W.}, \bibinfo{author}{Favaro, A.},
  \bibinfo{author}{Kinsler, P.} \& \bibinfo{author}{Boardman, A.}
\newblock \bibinfo{title}{A spacetime cloak, or a history editor}.
\newblock \emph{\bibinfo{journal}{J. Opt.}} \textbf{\bibinfo{volume}{13}},
  \bibinfo{pages}{024003} (\bibinfo{year}{2011}).

\bibitem{fridman2012demonstration}
\bibinfo{author}{Fridman, M.}, \bibinfo{author}{Farsi, A.},
  \bibinfo{author}{Okawachi, Y.} \& \bibinfo{author}{Gaeta, A.~L.}
\newblock \bibinfo{title}{Demonstration of temporal cloaking}.
\newblock \emph{\bibinfo{journal}{Nature}} \textbf{\bibinfo{volume}{481}},
  \bibinfo{pages}{62--65} (\bibinfo{year}{2012}).

\bibitem{lukens2013temporal}
\bibinfo{author}{Lukens, J.~M.}, \bibinfo{author}{Leaird, D.~E.} \&
  \bibinfo{author}{Weiner, A.~M.}
\newblock \bibinfo{title}{A temporal cloak at telecommunication data rate}.
\newblock \emph{\bibinfo{journal}{Nature}} \textbf{\bibinfo{volume}{498}},
  \bibinfo{pages}{205--208} (\bibinfo{year}{2013}).

\bibitem{chremmos2014temporal}
\bibinfo{author}{Chremmos, I.}
\newblock \bibinfo{title}{Temporal cloaking with accelerating wave packets}.
\newblock \emph{\bibinfo{journal}{Opt. Lett.}} \textbf{\bibinfo{volume}{39}},
  \bibinfo{pages}{4611--4614} (\bibinfo{year}{2014}).

\bibitem{philbin2008fiber}
\bibinfo{author}{Philbin, T.~G.} \emph{et~al.}
\newblock \bibinfo{title}{Fiber-optical analog of the event horizon}.
\newblock \emph{\bibinfo{journal}{Science}} \textbf{\bibinfo{volume}{319}},
  \bibinfo{pages}{1367--1370} (\bibinfo{year}{2008}).

\bibitem{westerberg2014experiment}
\bibinfo{author}{Westerberg, N.}, \bibinfo{author}{Cacciatori, S.},
  \bibinfo{author}{Belgiorno, F.}, \bibinfo{author}{Piazza, F.~D.} \&
  \bibinfo{author}{Faccio, D.}
\newblock \bibinfo{title}{Experimental quantum cosmology in time-dependent
  optical media}.
\newblock \emph{\bibinfo{journal}{New J. Phys.}} \textbf{\bibinfo{volume}{16}},
  \bibinfo{pages}{075003} (\bibinfo{year}{2014}).

\bibitem{graham1983jones}
\bibinfo{author}{Graham, E.~B.} \& \bibinfo{author}{Raab, R.~E.}
\newblock \bibinfo{title}{On the \uppercase{J}ones birefringence}.
\newblock \emph{\bibinfo{journal}{Proc. R. Soc. Lond. A}}
  \textbf{\bibinfo{volume}{390}}, \bibinfo{pages}{73--90}
  (\bibinfo{year}{1983}).

\bibitem{roth2000observation}
\bibinfo{author}{Roth, T.} \& \bibinfo{author}{Rikken, G. L. J.~A.}
\newblock \bibinfo{title}{Observation of magnetoelectric \uppercase{J}ones
  birefringence}.
\newblock \emph{\bibinfo{journal}{Phys. Rev. Lett.}}
  \textbf{\bibinfo{volume}{85}}, \bibinfo{pages}{4478} (\bibinfo{year}{2000}).

\bibitem{roth2000Magneto-electric}
\bibinfo{author}{Roth, T.} \& \bibinfo{author}{Rikken, G. L. J.~A.}
\newblock \bibinfo{title}{Magneto-electric \uppercase{J}ones birefringence: A
  bianisotropic effect}.
\newblock In \emph{\bibinfo{booktitle}{Bianisotropics 2000: 8th International
  Conference on Electromagnetics of Complex Media}}, \bibinfo{pages}{209--212}
  (\bibinfo{address}{Lisbon, Portugal}, \bibinfo{year}{2000}).

\bibitem{rizzo2003jones}
\bibinfo{author}{Rizzo, A.} \& \bibinfo{author}{Coriani, S.}
\newblock \bibinfo{title}{Jones birefringence in gases: ab initio electron
  correlated results for atoms and linear molecules}.
\newblock \emph{\bibinfo{journal}{J. Chem. Phys}}
  \textbf{\bibinfo{volume}{119}}, \bibinfo{pages}{11064--11079}
  (\bibinfo{year}{2003}).

\bibitem{mendonca2002time}
\bibinfo{author}{Mendonca, J.} \& \bibinfo{author}{Shukla, P.}
\newblock \bibinfo{title}{Time refraction and time reflection: two basic
  concepts}.
\newblock \emph{\bibinfo{journal}{Phys. Scripta}}
  \textbf{\bibinfo{volume}{65}}, \bibinfo{pages}{160} (\bibinfo{year}{2002}).

\bibitem{xiao2014reflection}
\bibinfo{author}{Xiao, Y.}, \bibinfo{author}{Maywar, D.~N.} \&
  \bibinfo{author}{Agrawal, G.~P.}
\newblock \bibinfo{title}{Reflection and transmission of electromagnetic waves
  at a temporal boundary}.
\newblock \emph{\bibinfo{journal}{Opt. Lett.}} \textbf{\bibinfo{volume}{39}},
  \bibinfo{pages}{574--577} (\bibinfo{year}{2014}).

\bibitem{wang2000gain}
\bibinfo{author}{Wang, L.~J.}, \bibinfo{author}{Kuzmich, A.} \&
  \bibinfo{author}{Dogariu, A.}
\newblock \bibinfo{title}{Gain-assisted superluminal light propagation}.
\newblock \emph{\bibinfo{journal}{Nature}} \textbf{\bibinfo{volume}{406}},
  \bibinfo{pages}{277--279} (\bibinfo{year}{2000}).

\bibitem{alexeev2002measurement}
\bibinfo{author}{Alexeev, I.}, \bibinfo{author}{Kim, K.~Y.} \&
  \bibinfo{author}{Milchberg, H.~M.}
\newblock \bibinfo{title}{Measurement of the superluminal group velocity of an
  ultrashort bessel beam pulse}.
\newblock \emph{\bibinfo{journal}{Phys. Rev. Lett.}}
  \textbf{\bibinfo{volume}{88}}, \bibinfo{pages}{073901}
  (\bibinfo{year}{2002}).

\bibitem{brunner2004direct}
\bibinfo{author}{Brunner, N.}, \bibinfo{author}{Scarani, V.},
  \bibinfo{author}{Wegm{\"u}ller, M.}, \bibinfo{author}{Legr{\'e}, M.} \&
  \bibinfo{author}{Gisin, N.}
\newblock \bibinfo{title}{Direct measurement of superluminal group velocity and
  signal velocity in an optical fiber}.
\newblock \emph{\bibinfo{journal}{Phys. Rev. Lett.}}
  \textbf{\bibinfo{volume}{93}}, \bibinfo{pages}{203902}
  (\bibinfo{year}{2004}).

\bibitem{bailly2010highly}
\bibinfo{author}{Bailly, G.}, \bibinfo{author}{Thon, R.} \&
  \bibinfo{author}{Robilliard, C.}
\newblock \bibinfo{title}{Highly sensitive frequency metrology for optical
  anisotropy measurements}.
\newblock \emph{\bibinfo{journal}{Rev. Sci. Instrum.}}
  \textbf{\bibinfo{volume}{81}}, \bibinfo{pages}{033105}
  (\bibinfo{year}{2010}).

\bibitem{robilliard2010towards}
\bibinfo{author}{Robilliard, C.} \& \bibinfo{author}{Bailly, G.}
\newblock \bibinfo{title}{Towards a first observation of magneto-electric
  directional anisotropy and linear birefringence in gases}.
\newblock \emph{\bibinfo{journal}{Can. J. Phys.}}
  \textbf{\bibinfo{volume}{89}}, \bibinfo{pages}{159--164}
  (\bibinfo{year}{2010}).

\bibitem{pelle2011magnetoelectric}
\bibinfo{author}{Pelle, B.}, \bibinfo{author}{Bitard, H.},
  \bibinfo{author}{Bailly, G.} \& \bibinfo{author}{Robilliard, C.}
\newblock \bibinfo{title}{Magnetoelectric directional nonreciprocity in
  gas-phase molecular \uppercase{N}itrogen}.
\newblock \emph{\bibinfo{journal}{Phys. Rev. Lett.}}
  \textbf{\bibinfo{volume}{106}}, \bibinfo{pages}{193003}
  (\bibinfo{year}{2011}).

\bibitem{durand2010shot}
\bibinfo{author}{Durand, M.}, \bibinfo{author}{Morville, J.} \&
  \bibinfo{author}{Romanini, D.}
\newblock \bibinfo{title}{Shot-noise-limited measurement of
  sub--parts-per-trillion birefringence phase shift in a high-finesse cavity}.
\newblock \emph{\bibinfo{journal}{Phys. Rev. A}} \textbf{\bibinfo{volume}{82}},
  \bibinfo{pages}{031803} (\bibinfo{year}{2010}).

\end{thebibliography}

\begin{thebibliography}{7}%
\makeatletter
\providecommand \@ifxundefined [1]{%
 \@ifx{#1\undefined}
}%
\providecommand \@ifnum [1]{%
 \ifnum #1\expandafter \@firstoftwo
 \else \expandafter \@secondoftwo
 \fi
}%
\providecommand \@ifx [1]{%
 \ifx #1\expandafter \@firstoftwo
 \else \expandafter \@secondoftwo
 \fi
}%
\providecommand \natexlab [1]{#1}%
\providecommand \enquote  [1]{``#1''}%
\providecommand \bibnamefont  [1]{#1}%
\providecommand \bibfnamefont [1]{#1}%
\providecommand \citenamefont [1]{#1}%
\providecommand \href@noop [0]{\@secondoftwo}%
\providecommand \href [0]{\begingroup \@sanitize@url \@href}%
\providecommand \@href[1]{\@@startlink{#1}\@@href}%
\providecommand \@@href[1]{\endgroup#1\@@endlink}%
\providecommand \@sanitize@url [0]{\catcode `\\12\catcode `\$12\catcode
  `\&12\catcode `\#12\catcode `\^12\catcode `\_12\catcode `\%12\relax}%
\providecommand \@@startlink[1]{}%
\providecommand \@@endlink[0]{}%
\providecommand \url  [0]{\begingroup\@sanitize@url \@url }%
\providecommand \@url [1]{\endgroup\@href {#1}{\urlprefix }}%
\providecommand \urlprefix  [0]{URL }%
\providecommand \Eprint [0]{\href }%
\providecommand \doibase [0]{http://dx.doi.org/}%
\providecommand \selectlanguage [0]{\@gobble}%
\providecommand \bibinfo  [0]{\@secondoftwo}%
\providecommand \bibfield  [0]{\@secondoftwo}%
\providecommand \translation [1]{[#1]}%
\providecommand \BibitemOpen [0]{}%
\providecommand \bibitemStop [0]{}%
\providecommand \bibitemNoStop [0]{.\EOS\space}%
\providecommand \EOS [0]{\spacefactor3000\relax}%
\providecommand \BibitemShut  [1]{\csname bibitem#1\endcsname}%
\let\auto@bib@innerbib\@empty
\bibitem [{\citenamefont {Carroll}\ \emph {et~al.}(1990)\citenamefont
  {Carroll}, \citenamefont {Field},\ and\ \citenamefont
  {Jackiw}}]{carroll1990limits2}%
  \BibitemOpen
  \bibfield  {author} {\bibinfo {author} {\bibfnamefont {S.~M.}\ \bibnamefont
  {Carroll}}, \bibinfo {author} {\bibfnamefont {G.~B.}\ \bibnamefont {Field}},
  \ and\ \bibinfo {author} {\bibfnamefont {R.}~\bibnamefont {Jackiw}},\
  }\href@noop {} {\bibfield  {journal} {\bibinfo  {journal} {Phys. Rev. D}\
  }\textbf {\bibinfo {volume} {41}},\ \bibinfo {pages} {1231} (\bibinfo {year}
  {1990})}\BibitemShut {NoStop}%
\bibitem [{\citenamefont {Itin}(2004)}]{itin2004carroll2}%
  \BibitemOpen
  \bibfield  {author} {\bibinfo {author} {\bibfnamefont {Y.}~\bibnamefont
  {Itin}},\ }\href@noop {} {\bibfield  {journal} {\bibinfo  {journal} {Phys.
  Rev. D}\ }\textbf {\bibinfo {volume} {70}},\ \bibinfo {pages} {025012}
  (\bibinfo {year} {2004})}\BibitemShut {NoStop}%
\bibitem [{\citenamefont {Itin}(2008)}]{itin2008wave2}%
  \BibitemOpen
  \bibfield  {author} {\bibinfo {author} {\bibfnamefont {Y.}~\bibnamefont
  {Itin}},\ }\href@noop {} {\bibfield  {journal} {\bibinfo  {journal} {Gen.
  Relativ. Gravit.}\ }\textbf {\bibinfo {volume} {40}},\ \bibinfo {pages}
  {1219} (\bibinfo {year} {2008})}\BibitemShut {NoStop}%
\bibitem [{\citenamefont {Diener}(1997)}]{diener1997energy}%
  \BibitemOpen
  \bibfield  {author} {\bibinfo {author} {\bibfnamefont {G.}~\bibnamefont
  {Diener}},\ }\href@noop {} {\bibfield  {journal} {\bibinfo  {journal} {Phys.
  Lett. A}\ }\textbf {\bibinfo {volume} {235}},\ \bibinfo {pages} {118}
  (\bibinfo {year} {1997})}\BibitemShut {NoStop}%
\bibitem [{\citenamefont {Brillouin}(1960)}]{brillouin1960wave}%
  \BibitemOpen
  \bibfield  {author} {\bibinfo {author} {\bibfnamefont {L.}~\bibnamefont
  {Brillouin}},\ }\href@noop {} {\emph {\bibinfo {title} {Wave propagation and
  group velocity}}}\ (\bibinfo  {publisher} {Academic Press},\ \bibinfo {year}
  {1960})\BibitemShut {NoStop}%
\bibitem [{\citenamefont {Milonni}(2004)}]{milonni2004fast}%
  \BibitemOpen
  \bibfield  {author} {\bibinfo {author} {\bibfnamefont {P.~W.}\ \bibnamefont
  {Milonni}},\ }\href@noop {} {\emph {\bibinfo {title} {Fast light, slow light
  and left-handed light}}}\ (\bibinfo  {publisher} {CRC Press},\ \bibinfo
  {year} {2004})\BibitemShut {NoStop}%
\bibitem [{\citenamefont {Zeidler}(2009)}]{zeidler2008quantum}%
  \BibitemOpen
  \bibfield  {author} {\bibinfo {author} {\bibfnamefont {E.}~\bibnamefont
  {Zeidler}},\ }\href@noop {} {\emph {\bibinfo {title} {Quantum Field Theory I:
  Basics in mathematics and physics}}}\ (\bibinfo  {publisher} {Springer},\
  \bibinfo {year} {2009})\BibitemShut {NoStop}%
\end{thebibliography}

\vspace{20pt}
\noindent
\textbf{\Large Acknowledgments}\\
\noindent
This work is supported by the National Science Foundation of China (Grant No.11475088 and 11275024) and by the Ministry of Science and Technology of China (2013YQ030595-3).

\vspace{20pt}
\noindent
\textbf{\Large Author contributions}\\
\noindent
M.L.G., R.Y.Z. and Q.Z. proposed the idea. R.Y.Z., Y.W.Z. and L.S.R. performed the theoretical derivation and analysis. W.W. provided suggestions about experimental design. M.L.G. and Q.Z. supervised the research. All authors contributed to the preparation of this manuscript.

\vspace{20pt}
\noindent
\textbf{\Large Additional information}\\
\noindent
\textbf{Competing financial interests} The authors declare no competing financial interests.

\onecolumngrid
\appendix
\clearpage


\begin{center}
\textbf{\large Supplemental Material: Time Circular Birefringence in Time-Dependent Magnetoelectric Media}
\end{center}
\setcounter{equation}{0}
\setcounter{figure}{0}
\setcounter{table}{0}
\setcounter{page}{1}
\makeatletter
\renewcommand{\theequation}{S\arabic{equation}}
\renewcommand{\thefigure}{S\arabic{figure}}
\renewcommand{\bibnumfmt}[1]{[S#1]}
\renewcommand{\citenumfont}[1]{S#1}

  In this supplementary information, we will give a further discussion about the time refraction and time reflection of TCB modes in arbitrary time dependent axion-type ME media, and will give the derivation of the wave front velocity $v_{\rm f}=v_1$ of TCB modes with the simplified dispersion relation $\omega_\pm=v_1 k\sqrt{(k\pm\beta)/k}$ in detail.

\subsection{Time refraction and time reflection of TCB modes in general conditions}
The two TCB modes in time-dependent axion-type ME media take the form $\bm{B}_{\pm}= T_\pm(t)\me^{\mi kz} \bm{\hat{U}}_\mp$, where the temporal parts $T_\pm(t)$ satisfy the following equation:
\begin{equation}\label{eigenequationS}
  \frac{\dif^2T_\pm}{\dif t^2}+\frac{\dif \ln\varepsilon}{\dif t}\frac{\dif T_\pm}{\dif t}+v^2 k(k\pm\mu\dot{\Theta})T_\pm=0.
\end{equation}
where $\varepsilon,\ \mu,\ \Theta$ are all functions of time in general. For simplicity, we demand all the parameters in Eq.~(\ref{eigenequationS}) are real. According to the Maxwell equations and the constitutive relations, the other three electromagnetic vectors of the corresponding TCB modes read
\begin{equation}\label{field relations_S}
  \bm{D}_\pm=\Theta\bm{B}_\pm\pm\frac{\varepsilon}{k}\dot{\bm{B}}_\pm,\qquad
  \bm{E}_\pm=\pm\frac{1}{k}\dot{\bm{B}}_\pm,\quad 
  \bm{H}_\pm=\frac{1}{\mu}\bm{B}_\pm\mp\frac{\Theta}{k}\dot{\bm{B}}_\pm.
\end{equation}
 Arbitrary two linearly independent solutions of linear Eq.~(\ref{eigenequationS}) can be regarded as the bases of its solution space. Supposing $g(t)$ and $h(t)$ are two independent real solutions of~(\ref{eigenequationS}), then other two solutions that are complex conjugates of each other can be constructed:
\begin{equation}\label{conjugate solution_S}
  T^1_\pm(t)=g(t)+\mi h(t)=\rho_\pm(t)\me^{\mi\psi_\pm(t)},\quad T^2_\pm(t)=g(t)-\mi h(t)=\rho_\pm(t)\me^{-\mi\psi_\pm(t)},
\end{equation}
where $\rho_\pm(t)$ and $\psi_\pm(t)$ are the amplitude and the polar angle of $T^1(t)_\pm$. $T^1_\pm(t)$ and $T^2_\pm(t)$ are also a set of bases of the solution space.  Therefore, $\bm{B}_\pm$ can always separate into two parts $\bm{B}_\pm=\bm{B}^1_\pm+\bm{B}^2_\pm$ with
\begin{equation}\label{two independent parts_S}
  \bm{B}^1_\pm=A^1_\pm \rho_\pm(t)\me^{\mi(kz+\psi_\pm(t))},\qquad \bm{B}^2_\pm=A^2_\pm \rho_\pm(t)\me^{\mi(kz-\psi_\pm(t))}.
\end{equation}

Consider a linearly polarized incident plane  wave $\bm{B}^{\rm in}=\bm{A}\me^{\mi(kz-\omega_0t)}=\sum_\pm (A/\sqrt{2})\me^{\mi(kz-\omega_0 t\pm\phi)}\hat{\bm{U}}_\mp$, as $t<t_0$, with $\omega_0=k/\sqrt{\varepsilon_0\mu_0}$ and $\bm{A}=\sum_\pm (A/\sqrt{2})\me^{\mp\mi\phi}\hat{\bm{U}}_\pm$, where $\phi$ is the polarized angle with respect to $x$ axis. After the wave passes through the time interface $t_0$ of the time wave plate, the wave becomes the superposition of the two TCB modes $\bm{B}=\sum_\pm\bm{B}_\pm$, and the two TCB modes can be further separate into two independent parts given in Eq.~(\ref{two independent parts_S}). In terms of the temporal boundary conditions, the coefficients of the two parts can be determined
\begin{equation}\label{coefficent_S}
  A_\pm^\sigma=\frac{A^{\rm in}_\pm\me^{\mi\left(\pm\phi-\omega_0 t_0+\delta^\sigma\psi_\pm(t_0)\right)}}{2\rho_\pm(t_0)\varepsilon_1(t_0)\dot{\psi}_\pm(t_0)}\delta^{\sigma}\Big[\big(\varepsilon_0\omega_0-\varepsilon_1(t_0)\omega^\sigma_\pm(t_0)\big)+\mi k\big(\Theta_1(t_0)-\Theta_0\big)\Big],\quad (\sigma=1,2),
\end{equation}
where $\omega^\sigma_\pm(t)=\mi\frac{\dif}{\dif t}\ln T^\sigma_\pm(t)=\frac{\dif}{\dif t}[\delta^\sigma\psi_\pm(t)+\mi\ln \rho_\pm(t)]$, and $A^{\rm in}_\pm$ is the amplitude of corresponding circularly polarized incident wave, for the case of linearly polarized incident wave,  $A^{\rm in}_\pm=A/\sqrt{2}\me^{\mp\mi\phi}$. If  $\dot{\Theta}_1(t)\equiv \beta/\mu_1>0$, and $\varepsilon_1,\ \mu_1$ are constant, Eq.~(\ref{coefficent_S}) reduces to the simplified expression given in  Eq.~(7) of the main text.

Because the lagrangian does not contain spatial coordinates explicitly, i.e. the system is invariant under spatial translation, according to Noether's theorem, the conservation of momentum of the system can be expressed as
\begin{equation}
  \frac{\partial}{\partial t}\left(\bm{D}\times\bm{B}\right)+\nabla\cdot\left[\frac{1}{2}\big(\bm{D}\cdot\bm{E}+\bm{B}\cdot\bm{H}\big)\matr{I}-\bm{D}\bm{E}-\bm{H}\bm{B}\right]=0,
\end{equation}
where $\bm{P}=\bm{D}\times\bm{B}$ is the apparent momentum density of electromagnetic fields, and $\matr{M}=\frac{1}{2}\big(\bm{D}\cdot\bm{E}+\bm{B}\cdot\bm{H}\big)\matr{I}-\bm{D}\bm{E}-\bm{H}\bm{B}$ is the Maxwell stress tensor. Since the TCB modes are transverse with respect to $\bm{k}$ and  their spatial parts merely vary with $z$, we have $\partial\bm{P}_\pm/\partial t=-\nabla\cdot\matr{M}_\pm=0$. Therefore, the momentum density $\bm{P}_\pm$ should be constant. 

\begin{figure}[t]
\includegraphics[height=0.47\columnwidth,clip]{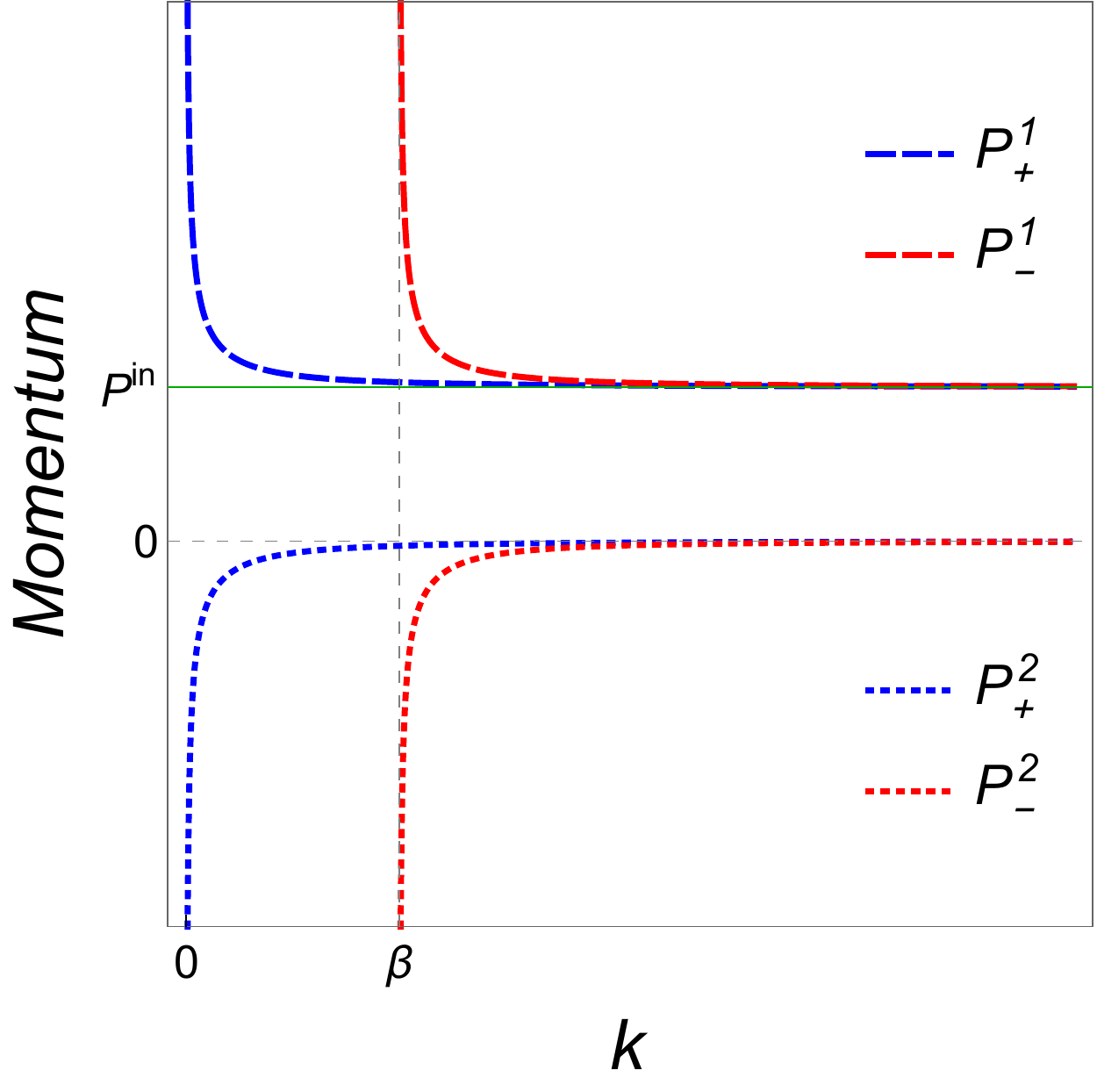}\caption{\label{fig_momentum}  Apparent momentum densities of the time refracted and time reflected parts of the two TCB modes versus $k$ for the simplified case  $\dot{\Theta}_1(t)\equiv \beta/\mu_1>0$, and $\varepsilon_1\equiv \varepsilon_0,\ \mu_1\equiv \mu_0$.}
\end{figure}

For $t>t_0$, the momentum density is $\bm{P}_\pm=\mathfrak{Re}(\bm{D}^1_\pm+\bm{D}^2_\pm)\times\mathfrak{Re}(\bm{B}^1_\pm+\bm{B}^2_\pm)=\bm{P}^1_\pm+\bm{P}^2_\pm+\bm{P}^{\rm cross}_\pm$, where $\bm{P}^1_\pm=\mathfrak{Re}(\bm{D}^1_\pm)\times\mathfrak{Re}(\bm{B}^1_\pm)$ and $\bm{P}^2_\pm=\mathfrak{Re}(\bm{D}^2_\pm)\times\mathfrak{Re}(\bm{B}^2_\pm)$ are the momentums of the two independent parts respectively, and $\bm{P}^{\rm cross}_\pm=\mathfrak{Re}(\bm{D}^1_\pm)\times\mathfrak{Re}(\bm{B}^2_\pm)+\mathfrak{Re}(\bm{D}^2_\pm)\times\mathfrak{Re}(\bm{B}^1_\pm)$ is the cross term. Substituting Eq.~(\ref{field relations_S}) and Eq.~(\ref{two independent parts_S}) into the momentum densities, we obtain the cross term is alway zero $\bm{P}^{\rm cross}_\pm\equiv 0$ and
$
  \bm{P}^\sigma_\pm=\delta^\sigma\frac{1}{2k}\varepsilon_1(t)\dot{\psi}_\pm(t)\rho_\pm(t)^2\left|C^\sigma_\pm\right|^2\hat{\bm{z}}\ (\sigma=1,2).
$
It is easy to check that $F=\varepsilon_1(t)\dot{\psi}_\pm(t)\rho_\pm(t)^2$ is a first integral of the ordinary differential equation (\ref{eigenequationS}). Actually, $F=\varepsilon_1(t)[\dot{g}(t)h(t)-g(t)\dot{h}(t)]$ according to Eq.~(\ref{conjugate solution_S}). Since $g(t)$ and $h(t)$ are both the solutions of (\ref{eigenequationS})
\begin{subequations}
\begin{align}
  \varepsilon_1\ddot{g}+\dot{\varepsilon}_1\dot{g}+\varepsilon_1 v^2 k(k\pm\mu\dot{\Theta})g & =0,\label{ODE1_S}\\
  \varepsilon_1\ddot{h}+\dot{\varepsilon}_1\dot{h}+\varepsilon_1 v^2 k(k\pm\mu\dot{\Theta})h & =0,\label{ODE2_S}
\end{align}
\end{subequations}
computing (\ref{ODE1_S})$\cdot h-$(\ref{ODE2_S})$\cdot g$ yields $\frac{\dif}{\dif t}F= 0$, so $F\equiv \varepsilon_1(t_0)\dot{\psi}_\pm(t_0)\rho_\pm(t_0)^2$. Then we obtain
\begin{equation}
  \bm{P}_\pm=\bm{P}^1_\pm+\bm{P}^2_\pm=\bm{P}^{\rm in}_\pm=\frac{\varepsilon_0\omega_0}{2k}|A^{\rm in}_\pm|^2\hat{\bm{z}},
\end{equation}
where the momentums of the two independent branches are respectively
\begin{equation}
  \bm{P}^\sigma_\pm=\delta^\sigma\frac{|A^{\rm in}_\pm|^2}{8k\varepsilon_1(t_0)\dot{\psi}_\pm(t_0)}\left[\Big(\varepsilon_0\omega_0-\delta^\sigma\varepsilon_1(t_0)\dot{\psi}_\pm(t_0)\Big)^2+\left(k\big(\Theta_1(t_0)-\Theta_0\big)-\varepsilon_1(t_0)\frac{\dot{\rho}_\pm(t_0)}{\rho_\pm(t_0)}\right)^2\right]\hat{\bm{z}},\quad (\sigma=1,2).
\end{equation}
 Fig.~\ref{fig_momentum} shows the momentums for the simplified case discussed in the main text. 
We can see that the momentums of the two parts are always in opposite directions. Supposing $\dot{\psi}_\pm(t_0)<0$, then $\bm{P}^1_\pm$ is always along the incident direction while $\bm{P}^2_\pm$ is along the inverse direction, and their vector sum always equals to the incident momentum. Therefore, $\bm{B}^1_\pm$ and $\bm{B}^2_\pm$ have clear physical meaning, i.e. the time refraction and the time reflection of the corresponding circularly polarized incident wave.

Similarly, we can calculate the Poynting vectors $\bm{S}_\pm=\mathfrak{Re}(\bm{E}^1_\pm+\bm{E}^2_\pm)\times\mathfrak{Re}(\bm{H}^1_\pm+\bm{H}^2_\pm)$ of the TCB modes. It can be demonstrated that the cross term of the time refracted and reflected parts also vanishes for each TCB mode: $\bm{S}^{\rm cross}_\pm=\mathfrak{Re}(\bm{E}^1_\pm)\times\mathfrak{Re}(\bm{H}^2_\pm)+\mathfrak{Re}(\bm{E}^2_\pm)\times\mathfrak{Re}(\bm{H}^1_\pm)=0$. Thus the total energy flow equals to the sum of the time refracted and reflected parts:
\begin{equation}
  \bm{S}_\pm=\bm{S}^1_\pm+\bm{S}^2_\pm,
\end{equation}
where
\begin{equation}
  \bm{S}^\sigma_\pm=\mathfrak{Re}(\bm{E}^\sigma_\pm)\times\mathfrak{Re}(\bm{H}^\sigma_\pm)=\frac{1}{\varepsilon_1(t)\mu_1(t)}\bm{P}^\sigma_\pm\quad (\sigma=1,2).
\end{equation}
Therefore, the Poynting vectors change with time in general, and $\bm{S}_\pm\neq\bm{S}^{\rm in}_\pm$ unless $\varepsilon_1\mu_1\equiv \varepsilon_0\mu_0$. In addition, the energy densities corresponding to the refracted and reflected parts are, respectively,
\begin{equation}
  W^\sigma_\pm=\frac{1}{2}\Big[\mathfrak{Re}(\bm{D}^\sigma_\pm)\cdot\mathfrak{Re}(\bm{E}^\sigma_\pm)+\mathfrak{Re}(\bm{B}^\sigma_\pm)\cdot\mathfrak{Re}(\bm{H}^\sigma_\pm)\Big]=\frac{|A^\sigma_\pm|^2}{4}\left[\frac{\varepsilon_1(t)}{k^2}\dot{\rho}_\pm(t)^2+\left(\frac{\varepsilon_1(t)\dot{\psi}_\pm(t)^2}{k^2}+\frac{1}{\mu_1(t)}\right)\rho_\pm(t)^2\right].
\end{equation} 
However, the cross term $W^{\rm cross}_\pm$ of the refracted and reflected parts does not equal to zero, so $W_\pm\neq W^1_\pm+W^2_\pm$.

It should be noted that the definition of time refracted and reflected parts is not unique, because we can arbitrarily choose the pair of independent real solutions $g(t)$ and $h(t)$ given in Eq.~(\ref{conjugate solution_S}). How to define the time refraction and reflection uniquely is still an open question in general situations. Nevertheless, to choose the pair of plane wave solutions $T_\pm^\sigma=\me^{\mi\delta^\sigma\omega_\pm t}$ as the time refraction and reflection seems quite reasonable in the simplified case discussed in the main text. In the situation of Gaussian pulse incidence, this choice ensures the defined refracted pulse and reflected pulse move in opposite directions with group velocities (see main text for details).  

For the simplified case, the Poynting vectors and energy densities take the form
\begin{subequations}\label{poynting and energy densities}
  \begin{align}
    \bm{S}^\sigma_\pm &=\langle\bm{S}^\sigma_\pm\rangle =-\delta^\sigma \frac{\omega_\pm}{2k\mu_1}|A^\sigma_\pm|^2\hat{\bm{z}},\\
    W^\sigma_\pm &=\langle W^\sigma_\pm\rangle =\frac{k\pm\beta/2}{2k\mu_1}|A^\sigma_\pm|^2.
  \end{align}
\end{subequations}
Therefore, the energy transport velocities of the two TCB modes are determined by
\begin{equation}\label{energy velocity_S}
    \bm{v}_{\rm E\pm}=\frac{\langle\bm{S}_\pm\rangle}{\langle W_\pm\rangle}=v_1\frac{\sqrt{k(k\pm\beta)}}{k\pm\beta/2}\bm{\hat{k}}.
\end{equation}
As we have discussed in the main text, the energy is not conserved generically in time-dependent systems. However, $\bm{S^\sigma_\pm}$ and $W^\sigma_\pm$ given in Eqs.~(\ref{poynting and energy densities}) are both invariant with time. The reason of this exceptional conservation is that the axion coupling in the lagrangian density  $\mathcal{L}_\Theta=\Theta(t)\bm{E}^\sigma_\pm\cdot\bm{B}^\sigma_\pm=0$ for the pair of TCB plane-wave modes. 

For ordinary spatial refraction and reflection, the conservation of energy leads to the equality of the incident light intensity $I^0=|\langle \bm{S}^{\rm in}\rangle|$ with the sum of the transmitted intensity $I^T=|\langle \bm{S}^1\rangle|$ and the reflected intensity $I^R=|\langle \bm{S}^2\rangle|$: $I^0=I^T+I^R$ (for one dimension), while this equality is false for time refraction and time reflection. If we follow the traditional definitions of transmissivity $T=I^T/I^0$ and reflectivity $R=I^R/I^0$, then $T+R\neq 1$ owing to energy non-conservation. However, we can introduce modified transmissivity and modified reflectivity respectively
\begin{equation}
  T=\frac{I^T}{I^{T+R}}=\frac{|\langle \bm{S}^1\rangle|}{|\langle \bm{S}^1\rangle|+|\langle \bm{S}^2\rangle|},\quad 
  R=\frac{I^R}{I^{T+R}}=\frac{|\langle \bm{S}^2\rangle|}{|\langle \bm{S}^1\rangle|+|\langle \bm{S}^2\rangle|}.
\end{equation}
Under these redefinitions, the normalization condition is satisfied: $T+R=1$ (see Fig.~2(a) in the main text).

\section{Front velocity of TCB modes} 
Ref.~\cite{carroll1990limits2,itin2004carroll2,*itin2008wave2} point out that the superluminal group velocities $v_{\rm g\pm}$ of CFJ modes indicate the violation of causality. However, our calculation in Eq.~(\ref{energy velocity_S}) shows that the energy transport velocity is lower than the speed of light in vacuum, although, the energy transport velocity defined in  Eq.~(\ref{energy velocity_S}) is more ``interpretive'' than measurable~\cite{diener1997energy}. By contrast, a more visualized definition is the front velocity which gives the speed of the wavefront of a step-function discontinuous wave and measures the speed of information propagation~\cite{brillouin1960wave,milonni2004fast}. In the following section, we will prove that the front velocity $v_{\rm f}\equiv v_1$ in the linearly varying axion-type media. 

Firstly, let's give the Fourier expansion with respect to wave vector $k$ for an arbitrary TM wave $\bm{B}(z,t)$ traveling along $z$ axis in linearly varying axion-type ME media:
\begin{equation}\label{Fourier expansion_S}
  \begin{split}
  \bm{B}(z,t) =\ \ &\int_{-\infty}^\infty\dif k \left( A^1_-(k)\me^{\mi\omega_-(k)t}+A^2_-(k)\me^{-\mi\omega_-(k)t}\right)\me^{\mi kz}\,\hat{\bm{U}}_+\\
   +&\int_{-\infty}^{\infty}\dif k \left( A^1_+(k)\me^{\mi\omega_+(k)t}+A^2_+(k)\me^{-\mi\omega_+(k)t}\right)\me^{\mi kz}\,\hat{\bm{U}}_-.
   \end{split}
\end{equation}
where the piecewise ``frequencies'' are
\begin{subequations}\label{piecewise frequency_S}
  \begin{gather}
  \omega_-(k)=\left\{
  \begin{aligned}
    &v_1k\sqrt{(k-\beta)/k} &  &k\in (-\infty,0)\cup(\beta,\infty)\\
    &\mi v_1 k\sqrt{(-k+\beta)/k} & &k\in [0,\beta]
  \end{aligned}
  \right.\\[5pt]
  \omega_+(k)=\left\{
  \begin{aligned}
    &v_1k\sqrt{(k+\beta)/k} &  &k\in (-\infty,-\beta)\cup(0,\infty)\\
    &-\mi v_1 k\sqrt{(-k-\beta)/k} & &k\in [-\beta,0]
  \end{aligned}
  \right.
  \end{gather}
\end{subequations}
 and they obey the relation
$\omega_+(-k)^*=-\omega_-(k)$. The corresponding piecewise ``phase velocities'' are $v_\pm(k)=\omega_\pm(k)/k$.
Because $\bm{B}$ is a real vector field: $\bm{B}(z,t)=\bm{B}(z,t)^*$, the Fourier coefficients are not independent:
\begin{equation}
  A_-^\sigma(-k)^*=A_+^\sigma(k),\qquad (\sigma=1,2).
\end{equation}
Therefore, Eq.~(\ref{Fourier expansion_S}) can be written as
\begin{equation}
  \bm{B}(z,t) = \int_{-\infty}^\infty\dif k \left( A^1_-(k)\me^{\mi\omega_-(k)t}+A^2_-(k)\me^{-\mi\omega_-(k)t}\right)\me^{\mi kz}\,\hat{\bm{U}}_+ +{\rm c.c.}
\end{equation}
We also can adopt another convention to define the piecewise ``frequencies'' and the piecewise ``phase velocities'':
\begin{equation}\label{second convention_S}
  \omega'_\pm(k)=\omega_\pm (k)^*,\quad v'_\pm(k)=v_\pm(k)^*.
\end{equation}
Here, we use symbols with prime, {\it e.g.} $\omega'_\pm(k)$ and $v'_\pm(k)$, to represent the second convention to differentiate from the first convention given in Eq.~(\ref{piecewise frequency_S}). The two conventions are obviously equivalent to each other, if the the following transform relations are satisfied:
\begin{equation}
\left\{
  \begin{aligned}
   & A'^\sigma_\pm(k)=A^\sigma_\pm{(k)}\ \ (\sigma=1,2), &  & \pm k\in (-\infty,-\beta)\cup(0,\infty)\\
   & A'^1_\pm(k)=A^2_\pm{(k)},\  A'^1_\pm(k)=A^2_\pm{(k)}, &  & \pm k\in [-\beta,0].
   \end{aligned}\right.
\end{equation}

Now we consider the time refraction and reflection of an incident pulse with two well-defined front edges $z=\pm a$ at the time interface $t_0=0$, i.e. 
$
\bm{B}(z,0)= B(z,0)\hat{\bm{U}}_+ +{\rm c.c.}=0$ for $|z|>a$, and $B(z,0)$ is supposed to be a smooth function. Without loss of the generality, we still choose $\varepsilon_1=\varepsilon_0$, $\mu_1=\mu_0$, $\Theta_1(0)=\Theta_0$, and assume that the incident pulse is merely superposed by the plane waves traveling towards the positive direction of $z$ axis:
\begin{equation}
  \bm{B}^{\rm in} (z,t)=\int_{-\infty}^\infty A^{\rm in}(k)\me^{\mi k(z-v_0 t)}\hat{\bm{U}}_+ +{\rm c.c.} \quad (t<0). 
\end{equation}
where $A^{\rm in}(k)$ is given by
\begin{equation}\label{fourier of B0_S}
  \begin{split}
  A^{\rm in}(k)=&\int_{-\infty}^\infty\dif z B(z,0)\me^{-\mi kz}=\int_{-a}^a\dif z B(z,0)\me^{-\mi kz}.
   \end{split}
\end{equation}
Since $B(z,0)$ is a smooth function with the compact support $[-a,a]$, according to the Paley–-Wiener–-Schwartz theorem  \cite{zeidler2008quantum}, $A^{\rm in}(\kappa)$ is analytic on the complex plane and satisfies
\begin{equation}\label{PWS thm_S}
  |A^{\rm in}(\kappa)|\leq \frac{C\,\me^{a|\mathfrak{Im}(\kappa)|}}{1+|\kappa|},
\end{equation}
for some constant $C$.

\begin{figure}[t]
\includegraphics[height=0.5\columnwidth,clip]{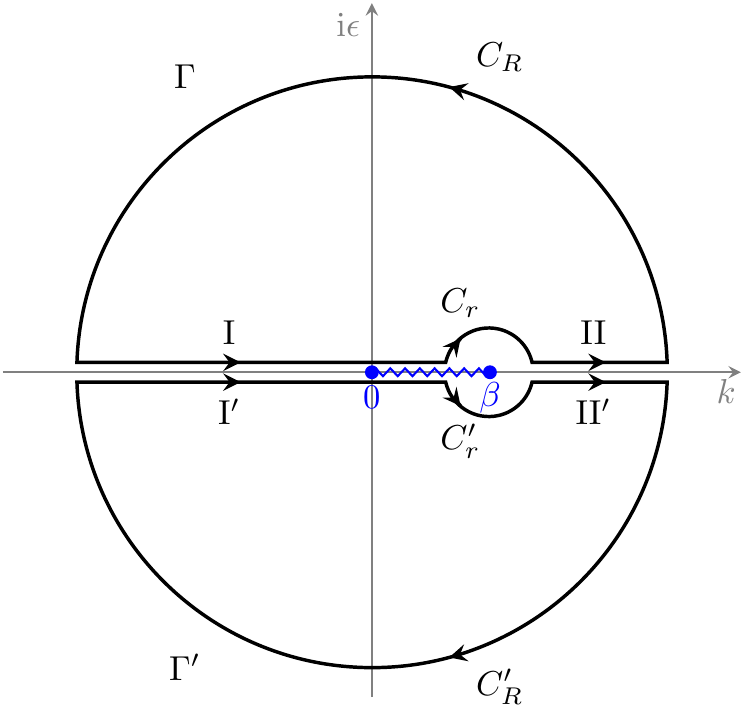}\caption{\label{fig_contour}  Contour $\Gamma$ and contour $\Gamma'$ for the two integrals given in Eq.~(\ref{contour integral_S}). Contour $\Gamma$ ($\Gamma'$) can separate into four parts: I, II, $C_r$, and $C_R$  ($\rm I'$, $\rm II'$, $C'_r$, and $C'_R$).  $k=0$ and $k=\beta$ on the real axis are two branch points of $u_\pm(\kappa)$ and the integrand, their connecting line is the branch cut.}
\end{figure}

According to the law of time refraction and time reflection for a particular wave vector given in Eq.~(6) and Eq.~(7) in the main text, we have
\begin{equation}\label{delay_S}
  \begin{split}
  \bm{B}(z,t)=&\int_{-\infty}^\infty \dif k\,\frac{1}{2}\left[\sum_{\sigma=1,2} \left(1-\delta^\sigma\frac{v_0}{v_-(k)}\right)\me^{\mi\delta^\sigma\omega_-(k)t}\right]A^{\rm in}(k)\,\me^{\mi kz}\,\hat{\bm{U}}_+ +{\rm c.c.}\\
  =& \sum_{\sigma=1,2}B^\sigma(z,t)\,\hat{\bm{U}}_+ +{\rm c.c.}\qquad (t>0) ,
  \end{split}
\end{equation}
where
\begin{equation}\label{propagator_S}
  B^\sigma(z,t)=\int_{-\infty}^\infty\dif k \frac{1}{2} \left(1-\delta^\sigma\frac{v_0}{v_-(k)}\right) A^{\rm in}(k)\,\me^{\mi k(z+\delta^\sigma v_-(k)t)} \quad (\sigma=1,2).
\end{equation} 
Note that the similar expression $B'^\sigma(z,t)$ is also valid for the second convention given in Eq.~(\ref{second convention_S}), and
\begin{equation}
  \sum_{\sigma=1,2}B^\sigma(z,t)=\sum_{\sigma=1,2}B'^\sigma(z,t).
\end{equation}
To extend the integral~(\ref{propagator_S}) to complex plane, we introduce two pairs of two-valued complex functions
\begin{equation}
   u_\pm(\kappa)=v_1 \sqrt{(\kappa\pm\beta)/\kappa},\qquad\Omega_\pm(\kappa)= \kappa\,u_\pm(\kappa)=v_1 \kappa\sqrt{(\kappa\pm\beta)/\kappa},
\end{equation}
with complex variable $\kappa=k+\mi \epsilon$. The two functions have two branch points $\kappa=0$, $\kappa=\mp\beta$ on real axis, and the line segment between the two points is the branch cut. Note that to choose the branch cut in this way follows the single-valued branch of square root function:
$
\sqrt{\kappa}:=   \sqrt{|\kappa|}\me^{\mi \arg(\kappa)/2}
$ with the convention $\arg(\kappa)\in (-\pi,\pi]$.
The limit from the upper half plane to real axis
\begin{equation}
  \begin{split}
  \lim_{\epsilon\rightarrow 0+}u_\pm(k+\mi\epsilon)=&\lim_{\epsilon\rightarrow 0+}v_1\sqrt{\left(1\pm\beta\frac{k}{k^2+\epsilon^2}\right)\mp\mi\frac{\beta\epsilon}{k^2+\epsilon^2}}=\lim_{\epsilon\rightarrow 0+}v_1\sqrt{\frac{k\pm\beta}{k}\mp\mi \epsilon}\\
  = &\left\{ \begin{aligned}
  & v_1\sqrt{(k\pm\beta)/k}  & & k\notin \text{Branch cut}\\
  & \mp\mi\sqrt{(-k\mp\beta)/k} & &k\in \text{Branch cut}
  \end{aligned}  \right.
    \end{split}
\end{equation}
is exactly the piecewise ``phase velocity'', and the limit of $\Omega_\pm(\kappa)$ is accordingly the piecewise ``frequency'' under the first convention, i.e.
\begin{equation}
  \lim_{\epsilon\rightarrow 0+}u_\pm(k+\mi\epsilon)=v_\pm(k),\qquad \lim_{\epsilon\rightarrow 0+}\Omega_\pm(k+\mi\epsilon)=\omega_\pm(k).
\end{equation}
On the contrary, the limits from the lower half plane to real axis gives the quantities under the second convention:
\begin{equation}
  \lim_{\epsilon\rightarrow 0-}u_\pm(k+\mi\epsilon)=v'_\pm(k),\qquad \lim_{\epsilon\rightarrow 0-}\Omega_\pm(k+\mi\epsilon)=\omega'_\pm(k).
\end{equation}
Therefore, $B^\sigma(z,t)$ defined in the first convention can be written as a complex integral
\begin{equation}\label{propagator2_S}
  \begin{split}
  B^\sigma(z,t)=&\int_{-\infty+\mi 0+}^{\infty+\mi 0+} \dif\kappa\, \frac{1}{2}\left(1-\delta^\sigma\frac{v_0}{u_-(\kappa)}\right)A^{\rm in}(\kappa)\me^{\mi \kappa(z+\delta^\sigma u_-(\kappa)t)}\\
  =&\left(\int_\mathrm{I}+\int_\mathrm{II}\right) \dif\kappa\, \tilde{B}^\sigma(k,t)\me^{\mi\kappa(z+\delta^\sigma v_1 t)},
  \end{split}
\end{equation}
and so do $B'^\sigma(z,t)$ defined in the second convention:
\begin{equation}\label{propagator3_S}
  \begin{split}
  B'^\sigma(z,t)=&\int_{-\infty+\mi 0-}^{\infty+\mi 0-} \dif\kappa\, \frac{1}{2}\left(1-\delta^\sigma\frac{v_0}{u_-(\kappa)}\right)A^{\rm in}(\kappa)\me^{\mi \kappa(z+\delta^\sigma u_-(\kappa)t)}\\
  =&\left(\int_\mathrm{I'}+\int_\mathrm{II'}\right) \dif\kappa\, \tilde{B}^\sigma(\kappa,t)\me^{\mi\kappa(z+\delta^\sigma v_1 t)},
  \end{split}
\end{equation}
where $\tilde{B}^\sigma(\kappa,t)$ is short for
\begin{equation}
  \tilde{B}^\sigma(\kappa,t)=\frac{1}{2}\left(1-\delta^\sigma\frac{v_0}{u_-(\kappa)}\right)A^{\rm in}(\kappa)\me^{\mi \kappa\delta^\sigma (u_-(\kappa)-v_1)t}.
\end{equation}
The integrals need to be separated into ${\rm I}$ and $\rm II$ (or ${\rm I}'$ and ${\rm II'}$) two parts because the branch point $\kappa=\beta$ is also a pole of the integrand. The integrand is analytic in the whole complex plane except the branch cut on the real axis, so the contour integrals with either the contour $\Gamma$ or the contour $\Gamma'$ shown in Fig.\ref{fig_contour} equal to zero
\begin{subequations}  \label{contour integral_S}
\begin{gather}\label{contour integral1_S}
  \oint_{\Gamma} \dif\kappa\, \tilde{B}^\sigma(\kappa,t)\me^{\mi\kappa(z+\delta^\sigma v_1 t)}=\left(\int_{\mathrm{I}}+\int_{\mathrm{II}}+\int_{C_r}+\int_{C_R}\right) \dif\kappa\,\tilde{B}^\sigma(\kappa,t)\me^{\mi\kappa(z+\delta^\sigma v_1 t)}=0,\\
\label{contour integral2_S}
 \oint_{\Gamma'} \dif\kappa\, \tilde{B}^\sigma(\kappa,t)\me^{\mi\kappa(z+\delta^\sigma v_1 t)}=\left(\int_{\mathrm{I'}}+\int_{\mathrm{II'}}+\int_{C'_r}+\int_{C'_R}\right) \dif\kappa\,\tilde{B}^\sigma(\kappa,t)\me^{\mi\kappa(z+\delta^\sigma v_1 t)}=0, 
\end{gather}
\end{subequations}
where $C_r$ is an infinitesimal semicircle with radius $r$ above the pole $k=\beta$,  $C_R$ is an infinite semicircle with radius $R$ in the upper half plane, and $C'_r$, $C'_R$ are their counterparts in the lower half plane. 

For the third term of Eq.~(\ref{contour integral1_S}),
\begin{equation}
  \begin{split}
  &\left|\int_{C_r}\dif\kappa\, \frac{1}{2}\left(1-\delta^\sigma\frac{v_0}{u_-(\kappa)}\right)A^{\rm in}(\kappa)\me^{\mi \kappa(z+\delta^\sigma u_-(\kappa)t)}\right|\\
=&\left|\int_{\pi}^0 r\me^{\mi\phi}\mi\dif\phi\,\frac{1}{2}\left(1-\delta^\sigma\sqrt{\frac{r\me^{\mi\phi}+\beta}{r\me^{\mi\phi}}}\right)A^{\rm in}(r\me^{\mi\phi}+\beta)\me^{\mi (r\me^{\mi\phi}+\beta)(z+\delta^\sigma u_-(r\me^{\mi\phi}+\beta)t)}\right|\\
\leq & \frac{\pi r}{2} \max \left[\left(1-\delta^\sigma\sqrt{\frac{r\me^{\mi\phi}+\beta}{r\me^{\mi\phi}}}\right)A^{\rm in}(r\me^{\mi\phi}+\beta)
\me^{\mi (r\me^{\mi\phi}+\beta)(z+\delta^\sigma u_-(r\me^{\mi\phi}+\beta)t)}\right]\\
\leq & \frac{\pi r}{2} \max \left|1-\delta^\sigma\sqrt{\frac{r\me^{\mi\phi}+\beta}{r\me^{\mi\phi}}}\right|
\max\left|A^{\rm in}(r\me^{\mi\phi}+\beta)\me^{\mi (r\me^{\mi\phi}+\beta)(z+\delta^\sigma u_-(r\me^{\mi\phi}+\beta)t)}\right|,
\end{split}
\end{equation}
where
\begin{equation}
  \max \left|1-\delta^\sigma\sqrt{\frac{r\me^{\mi\phi}+\beta}{r\me^{\mi\phi}}}\right|
  \leq \max\left( 1+\sqrt{\left|\frac{r\me^{\mi\phi}+\beta}{r\me^{\mi\phi}}\right|}\right)
  = 1+\sqrt{\frac{r+\beta}{r}},
\end{equation}
and $A^{\rm in}(r\me^{\mi\phi}+\beta)\me^{\mi (r\me^{\mi\phi}+\beta)(z+\delta^\sigma u_-(r\me^{\mi\phi}+\beta)t)}$ is bounded as $r\rightarrow 0$, therefore we have
\begin{equation}
  \lim_{r\rightarrow 0}\left|\int_{C_r}\dif\kappa\, \frac{1}{2}\left(1-\delta^\sigma\frac{v_0}{u_-(\kappa)}\right)A^{\rm in}(\kappa)\me^{\mi \kappa(z+\delta^\sigma u_-(\kappa)t)}\right|=0.
\end{equation}
So the third term has no contribution to the contour integral. 

For an integration $\int_{C_R}f(\kappa)\dif\kappa$ along the infinite semicircle $C_R$ in the upper half plane, if $|zf(z)|$ tends to zero uniformly when $|z|\rightarrow\infty$  both in the upper half plane and on the real axis, then the integral will vanish. We thus need to check the limit of $|\kappa\tilde{B}^\sigma(\kappa,t)\me^{\mi\kappa(z+\delta^\sigma v_1 t)}|$ as $|\kappa|\rightarrow \infty$ for calculating the fourth term  of Eq.~(\ref{contour integral1_S}). In light of  Eq.~(\ref{PWS thm_S}), we have the following inequality 
\begin{equation}
  \begin{split}
  \left|\kappa\tilde{B}(\kappa,t)\me^{\mi\kappa(z+\delta^\sigma v_1 t)}\right|=&|\kappa|\left|\frac{1}{2}\left(1-\delta^\sigma\frac{v_0}{u_-(\kappa)}\right)A^{\rm in}(\kappa)\me^{\mi \kappa\delta^\sigma (u_-(\kappa)-v_1)t}\me^{\mi\kappa(z+\delta^\sigma v_1 t)}\right|\\
  \leq & \frac{|\kappa|}{2}\left|\left(1-\delta^\sigma\frac{v_0}{u_-(\kappa)}\right)\me^{\mi \kappa\delta^\sigma (u_-(\kappa)-v_1)t}\right|\frac{C\,\me^{a \epsilon}}{1+|\kappa|}\me^{-\epsilon(z+\delta^\sigma v_1 t)}\\
  = & \frac{C}{2}\left|\left(1-\delta^\sigma\frac{v_0}{u_-(\kappa)}\right)\me^{\mi \kappa\delta^\sigma (u_-(\kappa)-v_1)t}\right|\frac{|\kappa|\,\me^{ -\epsilon(z-a+\delta^\sigma v_1 t)}}{1+|\kappa|}.
  \end{split}
\end{equation}
If $z-a+v_1t>z-a-v_1t>0$ for $t>0$, the exponential term tends to zero in the upper half plane. And on account of the limit $\lim_{|\kappa|\rightarrow\infty}u_-(\kappa)=v_1$, $|\kappa\tilde{B}^\sigma(\kappa,t)\me^{\mi\kappa(z+\delta^\sigma v_1 t)}|\rightarrow 0$ uniformly for $|\kappa|\rightarrow \infty$ both in the upper half plane and on the real axis. Thus the integral of the fourth term also vanishes when $R\rightarrow\infty$ as long as $z-a+v_1t>z-a-v_1t>0$. According to Eq.~(\ref{contour integral1_S}), the sum of the first two terms also should be zero as  $z-a-v_1t>0$  $(t>0)$, then we obtain
$  B^\sigma(z,t)=0\quad (\sigma=1,2)$, for $z-a- v_1t>0\ \text{and}\ t>0.$
Substituting this result into Eq.~(\ref{delay_S}) yields
\begin{equation}
  \bm{B}(z,t)=0,\quad \text{for}\ z>v_1 t+a\ \ (t>0).
\end{equation}
A similar analysis of the integral given in Eq.~(\ref{contour integral2_S}) leads to the result:
\begin{equation}
  \bm{B}(z,t)=0,\quad \text{for}\ z<-(v_1 t+a)\ \ (t>0).
\end{equation}
In other words, the wavefronts of both front and back edges can not propagate with a speed faster than $v_1$.

 On the other hand, if $|z|<v_1 t+a$, no matter $\Gamma$ or $\Gamma'$ is chosen to calculate $B^\sigma(z,t)$ (or $B'^\sigma(z,t)$), at least one of $B^1(z,t)$ and $B^2(z,t)$ would not be zero.  Therefore, $\bm{B}(z,t)\neq 0$ for $|z|<v_1 t+a\ (t>0)$.  In conclusion, the velocity of wave front is exactly $v_1$.

In the above discussion, we only concern the dispersion caused by $\dot{\Theta}=\beta$. In practice, the permittivity $\varepsilon_1$  and permeability $\mu_1$ have dispersion with $k$ in media, and accordingly $v_1=1/\sqrt{\varepsilon_1\mu_1}$ is also some function of $k$. The front velocity thus depends on the analyticity of $v_1(k)$ for actual materials and need to be further investigated. However, for the interaction between light and time-dependent true axion field in vacuum, or for the CFJ modes in Chern-Simons modified electrodynamics, the dispersion relation is precisely $\omega_\pm=c k\sqrt{(k\pm\beta)/k}$, so the front velocity $v_{\rm f}\equiv c$ and the causality will not be violated.

\end{document}